\documentclass[aps,prl,twocolumn,superscriptaddress,floatfix,longbibliography]{revtex4-2}

\usepackage[utf8]{inputenc}
\usepackage[T1]{fontenc}
\usepackage{amsmath, amssymb, bm}
\usepackage{graphicx}
\usepackage{xcolor}
\usepackage{hyperref}
\usepackage{cleveref}
\usepackage{booktabs}
\usepackage{algorithmic}
\usepackage{physics}

\hypersetup{
    colorlinks,
    linkcolor={blue!50!black},
    citecolor={blue!50!black},
    urlcolor={blue!80!black},
}

\newcommand{\vR}{\mathbf{R}}
\renewcommand{\vr}{\mathbf{r}}

\begin{document}

\title{Essentially No Energy Barrier Between\\Independent Fermionic Neural Quantum State Minima}

\author{David D. Dai}
\email{dddai@mit.edu}
\author{Marin Solja\v{c}i\'c}
\email{soljacic@mit.edu}
\affiliation{Department of Physics, Massachusetts Institute of Technology, Cambridge, Massachusetts 02139, USA}

\date{\today}

\begin{abstract}
Neural quantum states (NQS) have proven highly effective in representing quantum many-body wavefunctions, but their loss landscape remains poorly understood and debated.
Here, we demonstrate that the NQS loss landscape is more benign and similar to conventional deep learning than previously thought, exhibiting mode connectivity: independently trained NQS are connected by paths in parameter space with essentially no energy barrier.
To construct these paths, we develop GeoNEB, a path optimizer integrating efficient stochastic reconfiguration with the nudged elastic band method for constructing minimum energy paths. 
For the strongly interacting six-electron quantum dot modeled by a $1.6$M-parameter Psiformer, we find two independent minima with expected energy barrier $\sim10^{-5}$ times smaller than the system’s overall energy scale and $\sim10^{-3}$ times smaller than the linear path’s barrier.
The path respects physical symmetry in addition to achieving low energy, with the angular momentum remaining well quantized throughout.
Our work is the first to construct optimized paths between independently trained NQS, and it suggests that the NQS loss landscape may not be as pathological as once feared.
\end{abstract}

\maketitle

\section{Introduction}
Deep neural networks have introduced a new paradigm for solving the quantum many-body problem.
Neural quantum states (NQS)~\cite{carleo2017solving,nomura2017rbm,carleo2019netket,choo2019spincnn,sharir2020rnn,pfau2020ferminet,hermann2020paulinet,viteritti2023spinformer,vonglehn2023psiformer,hermann2023review,lange2024review} parameterize the wavefunction as a deep neural network $\Psi_\theta$ and train the parameters $\theta$ without external data by minimizing the expected energy of the ansatz~\cite{feynman1954og,mcmillan1965og,ceperley1977first,ceperley1980ground,tanatar1989twodeg,foulkes2001review,needs2010review}.
Recent architectures such as FermiNet~\cite{pfau2020ferminet,spencer2020better} and Psiformer~\cite{vonglehn2023psiformer} for continuum fermions can achieve chemical accuracy and surpass traditional methods on atomic and molecular systems~\cite{pfau2020ferminet,spencer2020better,vonglehn2023psiformer}.

It is currently unclear how transferable intuitions and patterns from conventional deep learning are to NQS~\cite{moss2025double}.
While first-order optimizers such as SGD~\cite{sutskever2013importance} or Adam~\cite{kingma2015adam} are standard and sufficient in natural language processing and computer vision~\cite{lecun2015deep}, NQS usually require sophisticated second-order optimizers~\cite{amari1998natural,sorella1998green,sorella2001lanczos,martens2015kfac,stokes2020natural} to achieve best results.
For example, training FermiNet and Psiformer using Adam instead of KFAC (Kronecker-factored approximate curvature) yields significantly worse energies~\cite{pfau2020ferminet, vonglehn2023psiformer}.
This uncertainty has led to opposing views regarding the NQS loss landscape's structure.
One view posits that with a sufficiently capable optimizer such as stochastic reconfiguration or KFAC, one can train and benefit from large and flexible models~\cite{pfau2020ferminet,vonglehn2023psiformer}.
Others argue that NQS optimization is hindered by an ``in general rugged and complicated loss landscape''~\cite{lange2025hybrid} and call for physically constrained ansatz instead~\cite{moss2025double}.

In this Letter, we investigate the NQS loss landscape through the lens of \textit{mode connectivity}.
In conventional deep learning, such as image classification, it is well-established that independently trained local minima are connected by continuous paths of near-constant low loss~\cite{garipov2018loss, draxler2018essentially}.
We extend this analysis to the quantum setting.
NQS optimization requires sampling from persistent Markov chains and preconditioning by the quantum geometric tensor, ruling out many path-finding methods from conventional deep learning.
To address these NQS-specific challenges, we introduce GeoNEB: Quantum Geometric Tensor-Informed Nudged Elastic Band.
It combines the nudged elastic band method~\cite{draxler2018essentially,jonsson1998neb,kolsbjerg2016autoneb} with exact natural gradient descent implemented efficiently using the kernel trick~\cite{chen2024empowering, rende2024simple}, explicitly respecting the quantum geometry of the parameter space.

For the strongly interacting six-electron quantum dot, we demonstrate two independently trained $1.6$M-parameter Psiformer models connected by a nonlinear path with energy barrier $7.2\times10^{-4}\text{ Ha}$ compared to the massive $40.817\text{ Ha}$ ground state energy and the $1.86\text{ Ha}$ barrier of the linear path.
We further show that the path respects physical symmetries, with angular momentum variance remaining small throughout the entire path.
The presented results were achieved for the first pair of minima trained, with minimal hyperparameter tuning.
The discovery of mode connectivity provides strong empirical evidence against the ``rugged landscape'' intuition, and it suggests that the NQS loss landscape is connected and reasonably accessible via stochastic reconfiguration.

\section{Neural Quantum States}

\begin{figure}[t!]
\centering
\includegraphics[width=\columnwidth]{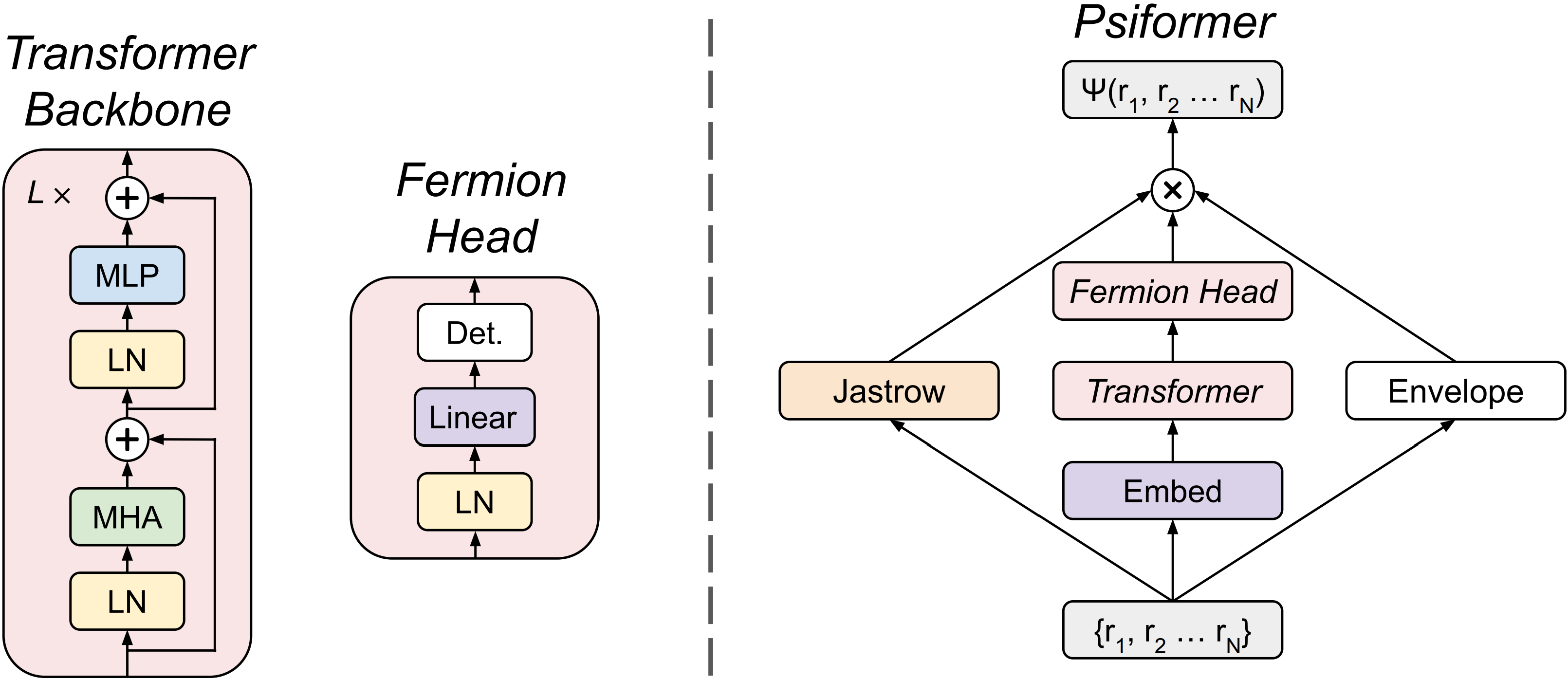}
\caption{Schematic of the Psiformer architecture. Electron coordinates are embedded and processed through transformer layers. A determinant enforces antisymmetry, while Jastrow and envelope factors handle cusps and boundary conditions.}
\label{fig:psiformer}
\end{figure}

We are concerned with $N$-electron Hamiltonians
\begin{align}
H = -\frac{1}{2}\sum_{i} \nabla_i^2 + \sum_{i < j} V_{\text{int}}(\vr_i-\vr_j) + \sum_{i} V_{\text{ext}}(\vr_i).
\end{align}
The ansatz $\Psi_\theta(\vR)$ must be antisymmetric under particle exchange, satisfy a cusp condition as two electrons approach each other, and obey the correct boundary conditions at infinity. We employ a modification of the Psiformer architecture (Fig.~\ref{fig:psiformer})~\cite{vonglehn2023psiformer}, an adaptation of the Transformer models so successful in natural language processing~\cite{vaswani2017attention} and computer vision~\cite{dosovitskiy2021vit}.

\textit{Transformer Backbone.}
The input configuration $\vR = \{\vr_1, \dots, \vr_N\}$ is embedded into feature vectors $\mathbf{h}_i^{0}$ using the raw Cartesian coordinates and ``soft radius'':
\begin{equation}
    \mathbf{h}_i^{0} = \mathbf{W}_{\text{embed}} \left( x_i, y_i, \sqrt{1 + r_i^2} \right).
\end{equation}
For the quantum dot, we found that using the soft radius $\sqrt{1+r_i^2}$ was essential.
Using the raw radius $r_i$ introduces an artificial cusp at the origin that causes divergent local energy, while removing radial information entirely deprives the network of an important feature.

Treating each electron as a token, these features pass through $L$ unmasked transformer layers.
Each layer applies multi-head self-attention (MHSA) followed by a GELU-activation multilayer perceptron (MLP) applying layer normalization (LN) before both sublayers~\cite{xiong2020prenorm,he2016prenorm}:
\begin{align}
    \tilde{\mathbf{h}}^{\ell} &= \mathbf{h}^{\ell-1} + \text{MHSA}\left(\text{LN}\left(\mathbf{h}^{\ell-1}\right)\right), \\
    \mathbf{h}^{\ell} &= \tilde{\mathbf{h}}^{\ell} + \text{MLP}\left(\text{LN}\left(\tilde{\mathbf{h}}^{\ell}\right)\right).
\end{align}

\textit{Jastrow and Envelope.}
The transformer output passes through an LN and then a linear layer to form the many-body orbitals used in the generalized Slater determinant.
We multiply the determinant by a Jastrow factor to enforce the electron-electron cusps and an envelope to enforce $r\rightarrow \infty$ boundary conditions:
\begin{equation}
\begin{aligned}
    \Phi&= \mathbf{W}_\text{orbital} \text{LN}\left(\textbf{h}^L\right) + \textbf{b}_\text{orbital},\\
    \Psi_\theta(\vR) &= \det\left[ \Phi \right] \times e^{-\sum_{i < j}\mathcal{J}(\vr_i-\vr_j)} \times \prod_{i} \mathcal{K}(\vr_i).\\
\end{aligned}
\end{equation}
The Jastrow and envelope forms for the quantum dot are
\begin{equation}
    \mathcal{J}(\vr) = \frac{\beta\alpha^2}{\alpha + r},\,\mathcal{K}(\vr) = e^{-\frac{r^2}{2}},
\end{equation}
where $\alpha$ is trainable and $\beta$ is fixed and matched to the spatial dimension and coupling strength.

\begin{figure}[t!]
\hrule
\vspace{2mm}
\textbf{Algorithm 1: GeoNEB (QGT-Informed NEB)}
\vspace{1mm}
\hrule
\begin{algorithmic}[1]
\REQUIRE $\theta_L, \theta_R$ (Minima), $N$ (Initial Pivots)
\STATE \textbf{Initialize:} $N$ pivots linearly between $\theta_L, \theta_R$
\WHILE{not converged}
    \FOR{$i = 1, \dots, N$}
        \STATE $\mathbf{R}_i \leftarrow$ MCMC sampling from $|\Psi_{\theta_i}|^2$
        \STATE $\mathbf{g}_{\text{nat}} \leftarrow \text{SR}(\theta_i, \mathbf{R}_i)$ \COMMENT{Eq. \ref{eq:exact_sr}}
        
        \STATE $\hat{\tau} \leftarrow \text{Normalize}(\theta_{i+1} - \theta_{i-1})$ 
        \STATE $\mathbf{g}_{\perp} \leftarrow \mathbf{g}_{\text{nat}} - (\mathbf{g}_{\text{nat}} \cdot \hat{\tau})\hat{\tau}$ 
        \STATE $\mathbf{g}_{\parallel} \leftarrow k ((2\theta_i - \theta_{i-1} - \theta_{i+1}) \cdot \hat{\tau})\hat{\tau}$ 
        
        \STATE $\theta_i \leftarrow \theta_i - \eta (\mathbf{g}_{\perp} + \mathbf{g}_{\parallel})$
    \ENDFOR
    
    \IF{iteration \% $T_{\text{insert}} == 0$}
        \STATE Insert pivots where loss is high
    \ENDIF
\ENDWHILE
\end{algorithmic}
\hrule
\caption{Pseudocode for GeoNEB. The curve consists of coupled NQS instances: each maintains its own Markov chains and local quantum geometric tensor but interacts with its neighbors to calculate the spring force and tangent.}
\label{alg:ngd_autoneb}
\end{figure}

\textit{Natural Gradient Optimization.}
Training minimizes $\bra{\Psi_\theta}H\ket{\Psi_\theta}/\bra{\Psi_\theta}\ket{\Psi_\theta} = \mathbb{E}_{\vR \sim |\Psi|^2} [E_L(\vR)]$, where the local energy is $E_L(\vR) = H\Psi(\vR)/\Psi(\vR)$.
To ameliorate the landscape's pathological curvature, we use stochastic reconfiguration (SR). While originally derived as the projection of imaginary time evolution onto the variational manifold, SR is in fact equivalent to natural gradient descent (NGD) using the quantum geometric tensor (QGT) to define the notion of distance.
For real wavefunctions, the QGT is equivalent to the Fisher information matrix (FIM) of the probability density $|\Psi_\theta|^2$.
Henceforth SR (QGT) and NGD (FIM) shall be interchangeable, $P\sim\order{10^6}$ shall denote the number of real parameters, and $M\sim\order{10^3}$ the batch size.

Remarkably, the exact SR/NGD update
\begin{equation}
    \Delta \theta = -\eta \left(\lambda \mathbf{I}_P + \mathbf{F}\right)^{-1} \mathbf{\nabla} E,   
\end{equation}
where $\mathbf{F}$ is the $P \times P$ QGT and $\lambda$ is the regularizing diagonal shift, can be performed without explicitly materializing the massive $\mathbf{F}$~\cite{chen2024empowering,rende2024simple}.
Defining the score $O_{i\mu} = \partial{\log \Psi(\vR_i)}/\partial{\theta_\mu}$, one can rewrite the update as~\footnote{We specialized to real wavefunctions here, but everything can be easily generalized to complex wavefunctions following~\cite{chen2024empowering,rende2024simple}.}
\begin{equation}\label{eq:exact_sr}
\begin{aligned}
    \mathbf{X}_{i\mu} &= \frac{1}{\sqrt{M}}\left(O_{i\mu} - \frac{1}{M}\sum_{j}O_{j\mu}\right),\\
    \mathbf{Y}_{i} &= \frac{2}{\sqrt{M}}\left(E_L(\vR_i) - \frac{1}{M}\sum_j E_L(\vR_j)\right),\\
    \Delta\theta &= -\eta \mathbf{X}^T \left(\lambda\mathbf{I}_M + \mathbf{X}\mathbf{X}^T\right)^{-1}\mathbf{Y}.
\end{aligned}
\end{equation}
The intractable $P \times P$ parameter-space inversion has been transformed into a relatively cheap $M \times M$ sample-space inversion. Compared to the KFAC that the original FermiNet and Psiformer implementations use~\cite{pfau2020ferminet,vonglehn2023psiformer}, exact SR/NGD avoids the Kronecker-factored structural approximation at the cost of some overhead~\cite{martens2015kfac}; however, this does not significantly reduce training speed because the Laplacian is so expensive regardless.

\begin{figure}[t!]
\centering
\includegraphics[width=\columnwidth]{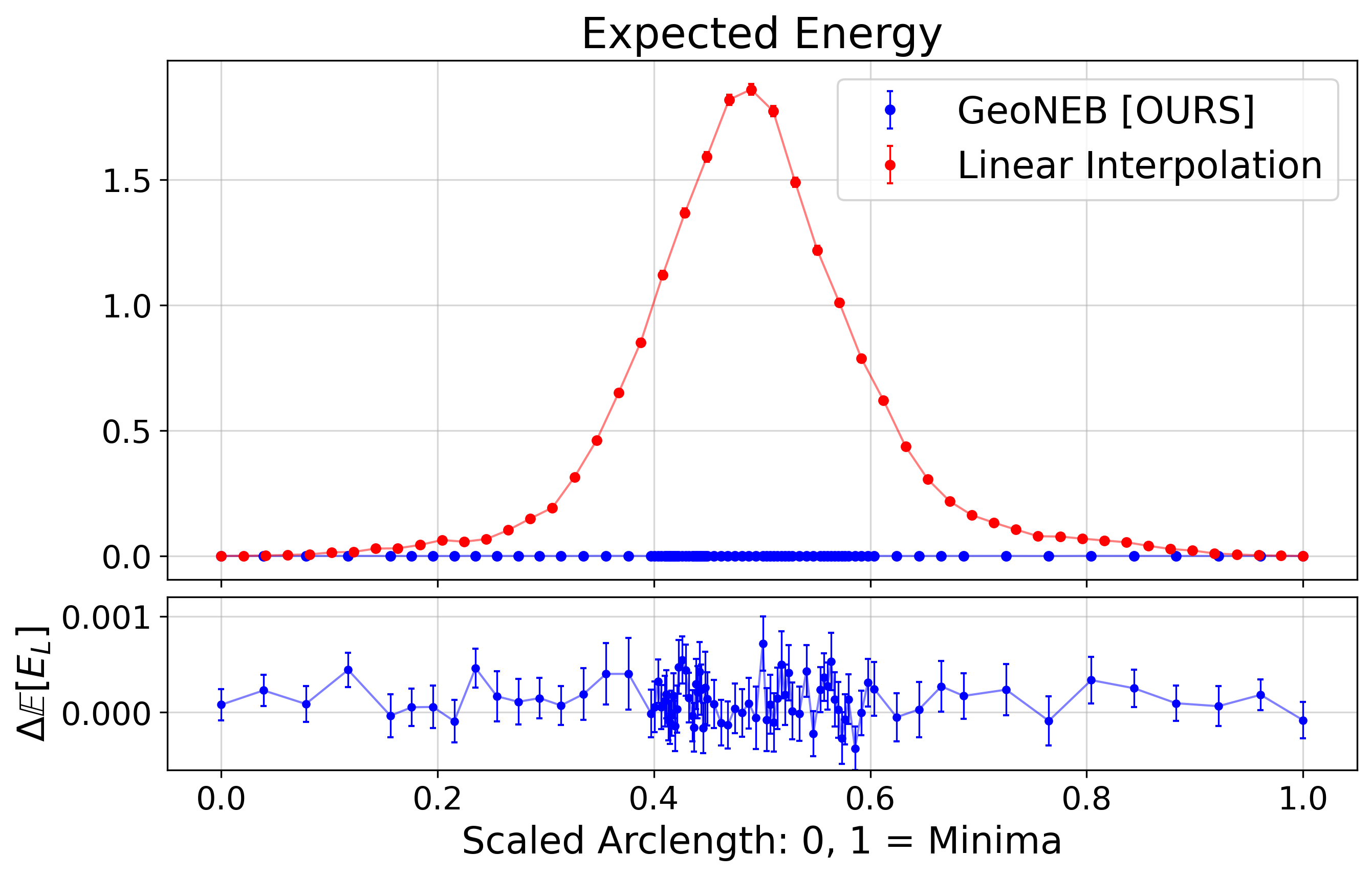}
\caption{\textit{Comparison of Energy Barriers.} GeoNEB finds a path with maximum barrier $7.2\times10^{-4}\text{ Ha}$, linear interpolation's barrier is $1.86\text{ Ha}$, and the ground state energy is $40.817\text{ Ha}$. The endpoints have Euclidean parameter norms $94.8$ and $95.1$, while their separation has norm $115.9$, confirming that the endpoints are far apart. Every fourth point is a trainable pivot; the others are inference-only.
}
\label{fig:energy}
\end{figure}

\section{Path Optimization Algorithm}

Finding a low-energy path between two minima $\theta_L$ and $\theta_R$ is a nontrivial optimization problem.
Naive linear interpolation $\theta(t) = (1-t)\theta_L + t\theta_R$ often traverses high-energy barriers.
We seek a curved path $\theta(t)$ that minimizes the maximum energy along the route.

\textit{Challenges in NQS.}
Finding such paths for NQS is uniquely challenging.
First, NQS training uses persistent Markov chains to sample from $|\Psi_\theta|^2$.
If the evaluation point $\theta(t)$ changes constantly at each iteration, such as if $t$ is sampled $\in[0, 1]$~\cite{garipov2018loss}, an expensive burn-in would be required at every iteration.
Second, SR requires a local QGT at the point of evaluation.
Supposing that we use a Bezier curve defined by the midpoint $\theta(0.5)$~\cite{garipov2018loss}, it is unclear how the loss gradient from $\theta(0.2)$ should be preconditioned, since $\theta(0.5)$ is the variational parameter for the curve, not $\theta(0.2)$, but those two parameters are different and have different local geometry.

\textit{GeoNEB Optimizer.}
We found the AutoNEB (Automated Nudged Elastic Band) algorithm~\cite{draxler2018essentially} to be most suitable for adaptation.
Instead of a continuous curve, it represents the curve as a discrete chain of ``pivots'' $\{p_0, \dots, p_{K+1}\}$, where $p_0=\theta_L$ and $p_{K+1}=\theta_R$.
The pivots interact via spring forces to maintain equal spacing but are otherwise independent NQS, each having its own persistent Markov chains and SR machinery.
Conceptually, we can view curve finding as ensemble training with an unusual form of regularization.

\begin{figure}[t!]
\centering
\includegraphics[width=\columnwidth]{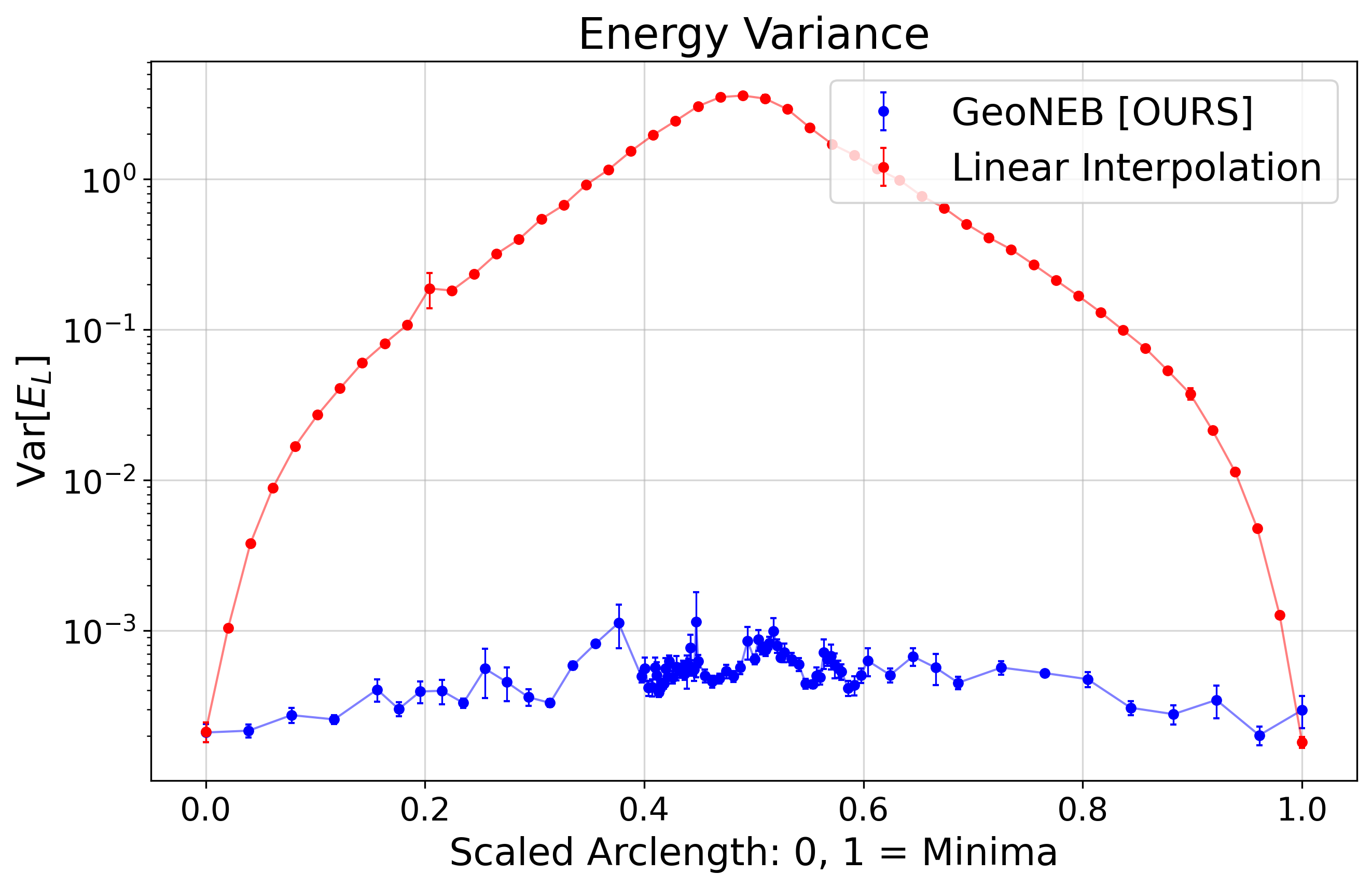}
\caption{\textit{Comparison of Energy Variance.} Plot uses log scale for visibility. GeoNEB's peak is $1.1\times10^{-3}\text{ Ha}^2$ while linear interpolation's peak is $3.6\text{ Ha}^2$. Every fourth point is a trainable pivot; the others are inference-only.
}
\label{fig:variance}
\end{figure}

\textit{Force Nudging.}
To prevent the pivots from sliding away from and undersampling the barrier, AutoNEB removes the component of the loss gradient parallel to the chain.
To prevent the spring force from overly straightening the chain, AutoNEB removes the perpendicular component of the spring force.
Letting $\tau_i$ be the chain's tangent at pivot $i$ and $k$ the spring strength, the update components for pivot $i$ are
\begin{align}
    \mathbf{g}^{\perp}_i &= \nabla_{\text{nat}} E(\theta_i) - \left(\nabla_{\text{nat}} E(\theta_i) \cdot \hat{\tau}_i\right)\hat{\tau}_i \\
    \mathbf{g}^{\parallel}_i &= k \left( (2\theta_i - \theta_{i-1} - \theta_{i+1}) \cdot \hat{\tau}_i \right) \hat{\tau}_i
\end{align}
Our contribution is to replace the raw loss gradient used originally with the natural (QGT-preconditioned) energy gradient.
We do not apply preconditioning to the spring force, because intuitively it should act directly in parameter space to maintain equal spacing there.

\textit{Dynamic Insertion.}
We initialize with a small number of pivots.
Periodically, we check the path resolution by evaluating an insertion criterion at the midpoints between pivots.
If a midpoint has worse criterion than the worst existing pivot, the midpoint is inserted as an additional pivot.
We used the energy variance as the criterion, because it usually scales linearly with energy error~\cite{fu2024variance} but exhibited better signal-to-noise ratio.
We do not redistribute pivots as done in~\cite{draxler2018essentially} because it would require burning in all Markov chains at every iteration.

\section{Results}

\textit{System.}
We selected the spin-polarized 2D quantum dot with six electrons in the strongly interacting regime (coupling $C=4$) as a testbed. The Hamiltonian is
\begin{equation}
H = -\frac{1}{2}\sum_{i} \nabla_i^2 + \sum_{i < j} \frac{C}{|\vr_i - \vr_j|} + \frac{1}{2}\sum_i \vr_i^2.
\end{equation}
Coupling $C=4$ is nontrivial because the electrons repel each other strongly, but do not repel so strongly that they become highly localized and thus easy to describe using a Slater determinant.
The modest system size enables very thorough inference of the connecting path to confirm that no loss barriers were missed.
The dot's circular symmetry yields the angular momentum as an independent metric of wavefunction quality.
At strong coupling, achieving a well-quantized angular momentum is nontrivial because the energetic penalty for symmetry breaking vanishes compared to the overall energy scale.

\begin{figure}[t!]
\centering
\includegraphics[width=\columnwidth]{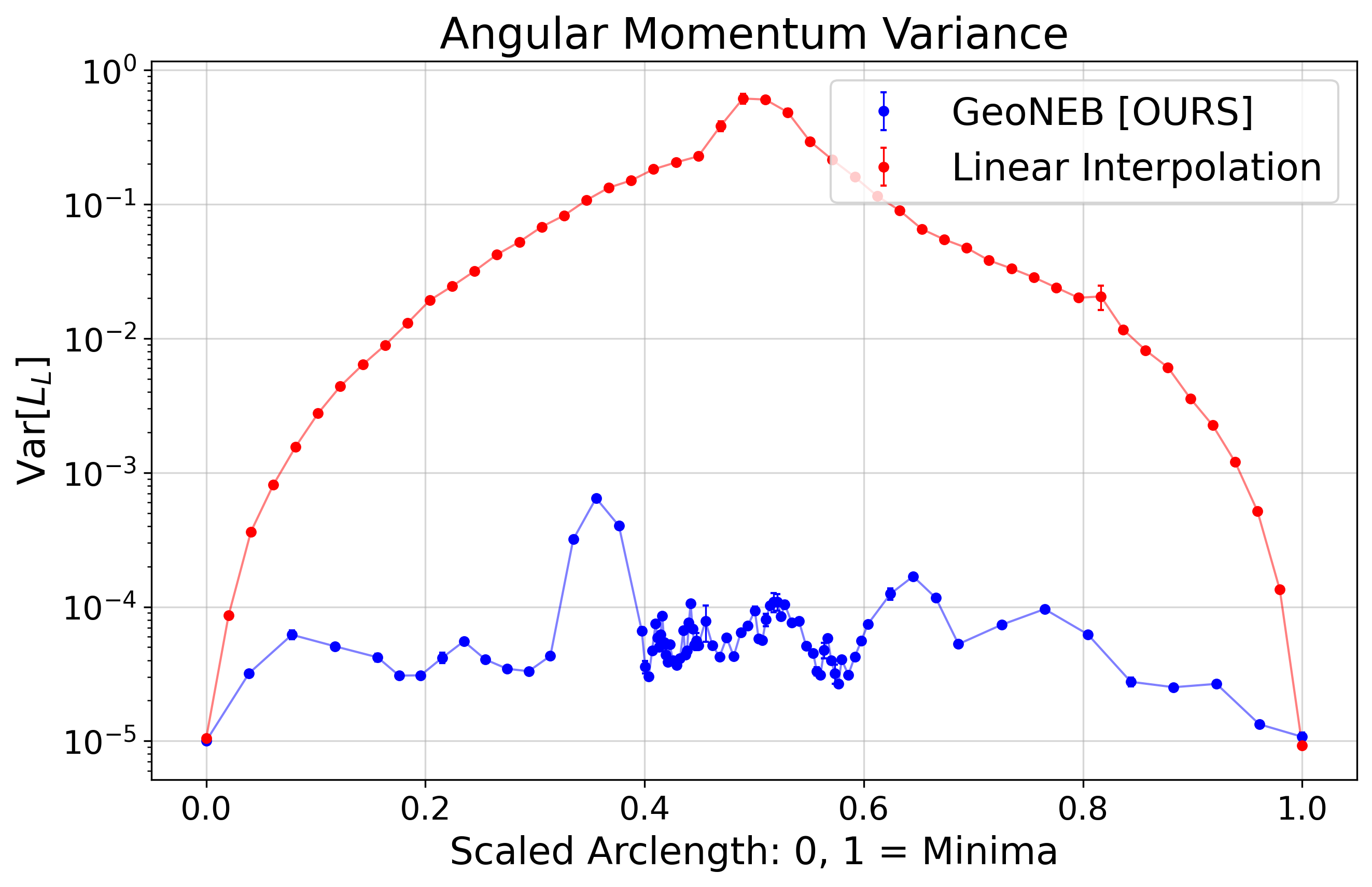}
\caption{\textit{Comparison of Angular Momentum Variance}. Plot uses log scale for visibility. GeoNEB's peak is $6.5\times10^{-4}\text{ a.u.}^2$ while linear interpolation's peak is $0.61\text{ a.u.}^2$. Every fourth point is a trainable pivot; the others are inference-only. The GeoNEB barrier is nontrivial, but in an absolute sense $6.5\times10^{-4}\text{ a.u.}^2$ is still good performance and the difference reflects the endpoints' excellence, not the path's shortcoming.}
\label{fig:angular}
\end{figure}

\textit{Setup.}
We trained two independent Psiformer models from random initializations, each having $1,586,182$ real parameters.
While conventional deep learning models scale to many billions of parameters, the high cost of Laplacian evaluation and Monte Carlo sampling unique to NQS limits practical models to about $\order{10^6}$ parameters.
We provide model configurations and training hyperparameters in the Supplementary Materials.
As shown in Table~\ref{tab:minima}, both converged to nearly identical energies $\sim 40.817\text{ Ha}$ and exceptionally low energy variances $\sim 3 \times 10^{-4}\text{ Ha}^2$.
The two minima are far from each other in weight space: the minima have Euclidean parameter norms of $94.8$ and $95.1$, while their separation has Euclidean norm $115.9$.

\textit{Expected Energy Barriers.}
We apply GeoNEB to find a low-energy path connecting the two minima.
Fig.~\ref{fig:energy} compares the energy profile along the GeoNEB-optimized path to that of the naive linear interpolation.
Every fourth point is an optimized pivot, while the remaining points were added at inference time to check for missed energy barriers.
The maximum rise along the GeoNEB path is $7.2\times10^{-4}\text{ Ha}$, while the maximum rise along the linear path is $\approx 1.86\text{ Ha}$, more than three orders of magnitude higher.

\begin{figure}[t!]
\centering
\includegraphics[width=\columnwidth]{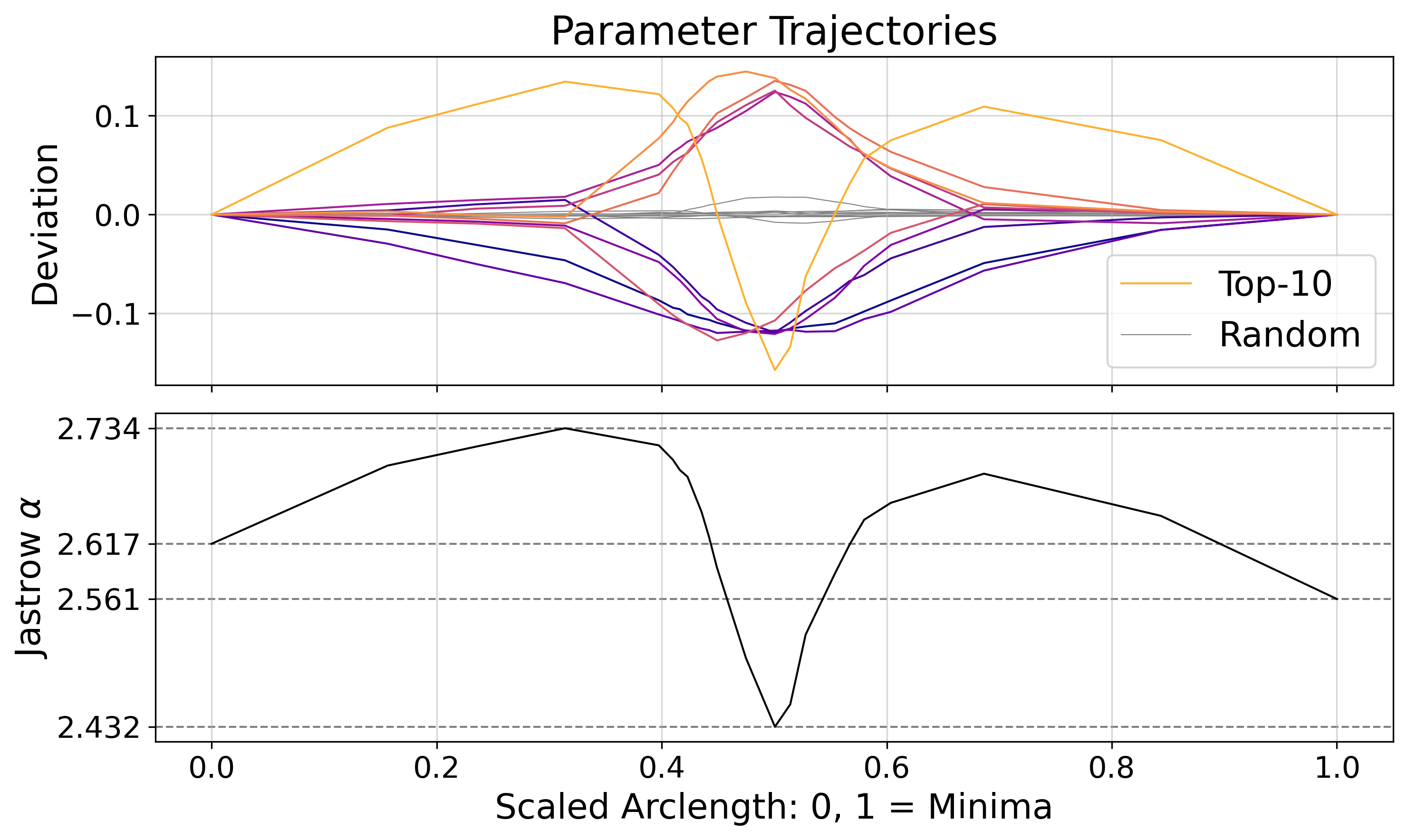}
\caption{\textit{GeoNEB Parameter Trajectories.} Top: Top-10 parameters with maximum absolute deviation from linear interpolation. Bottom: Jastrow $\alpha$ parameter along the path. Rapid variation in $\alpha$ between interpolation parameter $0.4$ and $0.6$ corresponds to the highest pivot concentration.}
\label{fig:params}
\end{figure}

\begin{table}[t]
\centering
\begin{tabular}{l|c|cc}
\toprule
Metric & Units & Minimum 1 & Minimum 2 \\
\midrule
$\langle H\rangle$ & $\text{Ha}$ & 40.81703(16) & 40.81687(19) \\
$\text{var}(H) = \langle H^2\rangle - \langle H \rangle^2$ & $\text{Ha}^2$ & 0.000210(28) & 0.000296(72) \\
$\text{var}(L) = \langle L^2\rangle - \langle L \rangle^2$ & $\text{a.u.}^2$ & 1.001(30)e-5 & 1.074(78)e-5 \\
\bottomrule
\end{tabular}
\caption{\textit{Results for Trained Minima}. All results are inferred using fresh test walkers never seen during training. The minima have Euclidean parameter norms $94.8$ and $95.1$, while their separation has norm $115.9$, confirming that the endpoints are far apart in parameter space.}
\label{tab:minima}
\end{table}

\textit{Operator Variance Barriers.}
To verify physical validity, we also examined the energy variance $\text{Var}[E_L]$ and the angular momentum variance $\text{Var}[L]$.
Fig.~\ref{fig:variance} shows that the energy variance along the GeoNEB path remains low, peaking at $1.1\times10^{-3}\text{ Ha}^2$ while the linear path spikes to $3.6\text{ Ha}^2$.
Fig.~\ref{fig:angular} shows that the angular momentum variance peak is $6.5\times10^{-4} \text{ a.u.}^2$, also about three orders of magnitude below linear interpolation.
We achieve these results despite the angular momentum never directly appearing in the loss function and circular symmetry not being explicitly enforced in the architecture.
While this is nontrivially larger than the endpoints, $6.5\times10^{-4} \text{ a.u.}^2$ is still very good performance, and this difference reflects the endpoints' excellence rather than the path's shortcoming.

\begin{figure*}[t!]
\centering
\includegraphics[width=0.98\textwidth]{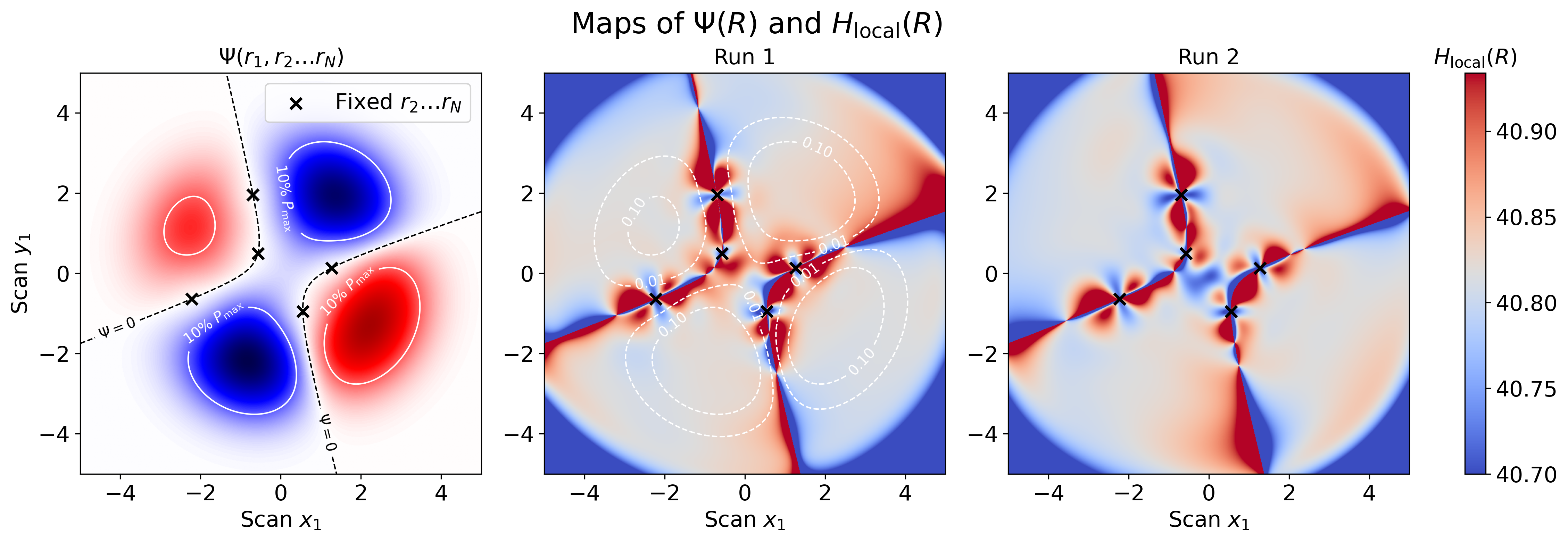}
\caption{\textit{Spatial Visualization of Trained Minima.} In all subplots, one electron is scanned around the plane while the others are clamped. Left: Image shows wavefunction amplitude with color representing sign and saturation representing magnitude; dashed lines mark nodal surfaces. There is no visible difference in $\Psi(\vR)$ for the two models. Middle/Right: Local energy maps for the two minima. While both minima have errors near the nodal surfaces, the fluctuations have different patterns, particularly in the center near $(0, 0)$ and upper nodal arm near $(-1, 3)$. The narrow plot range visually amplifies deviations.}
\label{fig:realspace}
\end{figure*}

\textit{Parameter Trajectories.}
Along the optimized path, most parameters follow smooth hump-shaped trajectories shown in Fig.~\ref{fig:params}.
However, the Jastrow factor's $\alpha$ parameter, which controls the radius over which the Jastrow is active, exhibits a sharp spike in the center of the path.
As a single scalar with disproportionate influence on the model output, $\alpha$'s rapid variation between interpolation parameter $0.4$ and $0.6$ forces dense sampling there.
Simultaneously, the good results obtained for a reasonable range of $\alpha$ along the path demonstrate that the rest of the model can compensate for variation in $\alpha$.

\textit{Local Energy Maps.}
Fig.~\ref{fig:realspace} visualizes the local energy $E_L(\vR)$ in real space and demonstrates that the models are legitimately different wavefunctions.
Both exhibit deviations near the nodal surfaces where $\Psi=0$, but the precise local structure of these errors exhibits some differences.
The performance degradation near the nodal surface is both excusable and difficult to fix, since the nodal surface both contributes very little to expectation values and receives very little training signal.

\section{Discussion}
We report the first observation of mode connectivity in neural quantum states using optimized paths.
Because our path achieves quantized angular momentum in addition to low energy, the loss landscape is connected not only from the perspective of pure energy minimization, but also holistic physical validity.
We achieved our results for the first pair of minima attempted and with minimal tuning of curve-optimization hyperparameters, indicating mode connectivity's robustness.
While there are real differences between the NQS loss landscape and conventional deep learning ones, our work suggests that stochastic reconfiguration resolves most pathologies and yields a reasonably benign landscape exhibiting similar virtues such as mode connectivity.

Our work also provides insight into the critical role of the Jastrow in NQS optimization.
The Jastrow $\alpha$ parameter's rapid variation near the path center suggests that such ``keystone parameters'' may complicate the landscape geometry and harm optimization.
To our knowledge, this type of parameter is unique to NQS architectures.
Because a nontrivial range of $\alpha$ values lie on the low-energy manifold, and even the directly trained endpoints have different $\alpha$, freezing the Jastrow into a hyperparameter could improve optimization with negligible accuracy loss.
Currently, we also do not know if the rapid variation is an optimization artifact or a genuine unavoidable feature of the loss landscape.
Based on our experience training NQS, we suspect that it could be an artifact caused by the more aggressive learning rate schedule used during curve optimization, although we defer detailed investigation to future work.

The transfer of mode connectivity to NQS from conventional deep learning was far from guaranteed despite similar high-level intuition.
Much of the existing theory of mode connectivity relies heavily on structural assumptions specific to supervised classification and regression.
For example, the noise stability theory assumes a static dataset and a per-sample loss that depends solely on network output and is convex in it~\cite{kuditipudi2019noise}.
NQS unsupervised energy minimization violates both assumptions: the data distribution is defined by the model itself, and the per-sample local energy loss depends on the network Laplacian.
For deep learning, our work expands our understanding of mode connectivity by observing it in a task with very different characteristics from the typical supervised classification.
For physics, our work suggests that the NQS loss landscape is not as pathological as once feared, and that overparameterization does not prevent optimization and is potentially even beneficial.

\textit{Note on Concurrent Work.}
A concurrent preprint~\cite{hernandes2025mode} studied linear interpolations rather than optimized curves connecting minima.
Consequently, they found high barriers between independent models, identifying low-barrier paths only between models related by fine-tuning or transfer learning.
Our work is the first to observe mode connectivity in the strict sense---an optimized path connecting independently initialized and trained models---as defined in the conventional deep learning literature.
We appreciate their complementary perspective on transfer learning and refer the reader to their work for insights into that domain.

\section*{Acknowledgments}
DDD was supported by the National Science Foundation Graduate Research Fellowship under Grant No. 2141064.
The authors acknowledge the MIT SuperCloud and Lincoln Laboratory Supercomputing Center for providing high-performance computing resources.

\bibliography{references}

\begin{thebibliography}{43}%
\makeatletter
\providecommand \@ifxundefined [1]{%
 \@ifx{#1\undefined}
}%
\providecommand \@ifnum [1]{%
 \ifnum #1\expandafter \@firstoftwo
 \else \expandafter \@secondoftwo
 \fi
}%
\providecommand \@ifx [1]{%
 \ifx #1\expandafter \@firstoftwo
 \else \expandafter \@secondoftwo
 \fi
}%
\providecommand \natexlab [1]{#1}%
\providecommand \enquote  [1]{``#1''}%
\providecommand \bibnamefont  [1]{#1}%
\providecommand \bibfnamefont [1]{#1}%
\providecommand \citenamefont [1]{#1}%
\providecommand \href@noop [0]{\@secondoftwo}%
\providecommand \href [0]{\begingroup \@sanitize@url \@href}%
\providecommand \@href[1]{\@@startlink{#1}\@@href}%
\providecommand \@@href[1]{\endgroup#1\@@endlink}%
\providecommand \@sanitize@url [0]{\catcode `\\12\catcode `\$12\catcode `\&12\catcode `\#12\catcode `\^12\catcode `\_12\catcode `\%12\relax}%
\providecommand \@@startlink[1]{}%
\providecommand \@@endlink[0]{}%
\providecommand \url  [0]{\begingroup\@sanitize@url \@url }%
\providecommand \@url [1]{\endgroup\@href {#1}{\urlprefix }}%
\providecommand \urlprefix  [0]{URL }%
\providecommand \Eprint [0]{\href }%
\providecommand \doibase [0]{https://doi.org/}%
\providecommand \selectlanguage [0]{\@gobble}%
\providecommand \bibinfo  [0]{\@secondoftwo}%
\providecommand \bibfield  [0]{\@secondoftwo}%
\providecommand \translation [1]{[#1]}%
\providecommand \BibitemOpen [0]{}%
\providecommand \bibitemStop [0]{}%
\providecommand \bibitemNoStop [0]{.\EOS\space}%
\providecommand \EOS [0]{\spacefactor3000\relax}%
\providecommand \BibitemShut  [1]{\csname bibitem#1\endcsname}%
\let\auto@bib@innerbib\@empty
\bibitem [{\citenamefont {Carleo}\ and\ \citenamefont {Troyer}(2017)}]{carleo2017solving}%
  \BibitemOpen
  \bibfield  {author} {\bibinfo {author} {\bibfnamefont {G.}~\bibnamefont {Carleo}}\ and\ \bibinfo {author} {\bibfnamefont {M.}~\bibnamefont {Troyer}},\ }\bibfield  {title} {\bibinfo {title} {Solving the quantum many-body problem with artificial neural networks},\ }\href {https://doi.org/10.1126/science.aag2302} {\bibfield  {journal} {\bibinfo  {journal} {Science}\ }\textbf {\bibinfo {volume} {355}},\ \bibinfo {pages} {602} (\bibinfo {year} {2017})},\ \Eprint {https://arxiv.org/abs/https://www.science.org/doi/pdf/10.1126/science.aag2302} {https://www.science.org/doi/pdf/10.1126/science.aag2302} \BibitemShut {NoStop}%
\bibitem [{\citenamefont {Nomura}\ \emph {et~al.}(2017)\citenamefont {Nomura}, \citenamefont {Darmawan}, \citenamefont {Yamaji},\ and\ \citenamefont {Imada}}]{nomura2017rbm}%
  \BibitemOpen
  \bibfield  {author} {\bibinfo {author} {\bibfnamefont {Y.}~\bibnamefont {Nomura}}, \bibinfo {author} {\bibfnamefont {A.~S.}\ \bibnamefont {Darmawan}}, \bibinfo {author} {\bibfnamefont {Y.}~\bibnamefont {Yamaji}},\ and\ \bibinfo {author} {\bibfnamefont {M.}~\bibnamefont {Imada}},\ }\bibfield  {title} {\bibinfo {title} {Restricted {B}oltzmann machine learning for solving strongly correlated quantum systems},\ }\href {https://doi.org/10.1103/PhysRevB.96.205152} {\bibfield  {journal} {\bibinfo  {journal} {Phys. Rev. B}\ }\textbf {\bibinfo {volume} {96}},\ \bibinfo {pages} {205152} (\bibinfo {year} {2017})}\BibitemShut {NoStop}%
\bibitem [{\citenamefont {Carleo}\ \emph {et~al.}(2019)\citenamefont {Carleo}, \citenamefont {Choo}, \citenamefont {Hofmann}, \citenamefont {Smith}, \citenamefont {Westerhout}, \citenamefont {Alet}, \citenamefont {Davis}, \citenamefont {Efthymiou}, \citenamefont {Glasser}, \citenamefont {Lin}, \citenamefont {Mauri}, \citenamefont {Mazzola}, \citenamefont {Mendl}, \citenamefont {{van Nieuwenburg}}, \citenamefont {O’Reilly}, \citenamefont {Théveniaut}, \citenamefont {Torlai}, \citenamefont {Vicentini},\ and\ \citenamefont {Wietek}}]{carleo2019netket}%
  \BibitemOpen
  \bibfield  {author} {\bibinfo {author} {\bibfnamefont {G.}~\bibnamefont {Carleo}}, \bibinfo {author} {\bibfnamefont {K.}~\bibnamefont {Choo}}, \bibinfo {author} {\bibfnamefont {D.}~\bibnamefont {Hofmann}}, \bibinfo {author} {\bibfnamefont {J.~E.}\ \bibnamefont {Smith}}, \bibinfo {author} {\bibfnamefont {T.}~\bibnamefont {Westerhout}}, \bibinfo {author} {\bibfnamefont {F.}~\bibnamefont {Alet}}, \bibinfo {author} {\bibfnamefont {E.~J.}\ \bibnamefont {Davis}}, \bibinfo {author} {\bibfnamefont {S.}~\bibnamefont {Efthymiou}}, \bibinfo {author} {\bibfnamefont {I.}~\bibnamefont {Glasser}}, \bibinfo {author} {\bibfnamefont {S.-H.}\ \bibnamefont {Lin}}, \bibinfo {author} {\bibfnamefont {M.}~\bibnamefont {Mauri}}, \bibinfo {author} {\bibfnamefont {G.}~\bibnamefont {Mazzola}}, \bibinfo {author} {\bibfnamefont {C.~B.}\ \bibnamefont {Mendl}}, \bibinfo {author} {\bibfnamefont {E.}~\bibnamefont {{van Nieuwenburg}}}, \bibinfo {author} {\bibfnamefont {O.}~\bibnamefont {O’Reilly}}, \bibinfo {author} {\bibfnamefont
  {H.}~\bibnamefont {Théveniaut}}, \bibinfo {author} {\bibfnamefont {G.}~\bibnamefont {Torlai}}, \bibinfo {author} {\bibfnamefont {F.}~\bibnamefont {Vicentini}},\ and\ \bibinfo {author} {\bibfnamefont {A.}~\bibnamefont {Wietek}},\ }\bibfield  {title} {\bibinfo {title} {{N}et{K}et: {A} machine learning toolkit for many-body quantum systems},\ }\href {https://doi.org/https://doi.org/10.1016/j.softx.2019.100311} {\bibfield  {journal} {\bibinfo  {journal} {SoftwareX}\ }\textbf {\bibinfo {volume} {10}},\ \bibinfo {pages} {100311} (\bibinfo {year} {2019})}\BibitemShut {NoStop}%
\bibitem [{\citenamefont {Choo}\ \emph {et~al.}(2019)\citenamefont {Choo}, \citenamefont {Neupert},\ and\ \citenamefont {Carleo}}]{choo2019spincnn}%
  \BibitemOpen
  \bibfield  {author} {\bibinfo {author} {\bibfnamefont {K.}~\bibnamefont {Choo}}, \bibinfo {author} {\bibfnamefont {T.}~\bibnamefont {Neupert}},\ and\ \bibinfo {author} {\bibfnamefont {G.}~\bibnamefont {Carleo}},\ }\bibfield  {title} {\bibinfo {title} {Two-dimensional frustrated ${J}_{1}\text{\ensuremath{-}}{J}_{2}$ model studied with neural network quantum states},\ }\href {https://doi.org/10.1103/PhysRevB.100.125124} {\bibfield  {journal} {\bibinfo  {journal} {Phys. Rev. B}\ }\textbf {\bibinfo {volume} {100}},\ \bibinfo {pages} {125124} (\bibinfo {year} {2019})}\BibitemShut {NoStop}%
\bibitem [{\citenamefont {Sharir}\ \emph {et~al.}(2020)\citenamefont {Sharir}, \citenamefont {Levine}, \citenamefont {Wies}, \citenamefont {Carleo},\ and\ \citenamefont {Shashua}}]{sharir2020rnn}%
  \BibitemOpen
  \bibfield  {author} {\bibinfo {author} {\bibfnamefont {O.}~\bibnamefont {Sharir}}, \bibinfo {author} {\bibfnamefont {Y.}~\bibnamefont {Levine}}, \bibinfo {author} {\bibfnamefont {N.}~\bibnamefont {Wies}}, \bibinfo {author} {\bibfnamefont {G.}~\bibnamefont {Carleo}},\ and\ \bibinfo {author} {\bibfnamefont {A.}~\bibnamefont {Shashua}},\ }\bibfield  {title} {\bibinfo {title} {Deep autoregressive models for the efficient variational simulation of many-body quantum systems},\ }\href {https://doi.org/10.1103/PhysRevLett.124.020503} {\bibfield  {journal} {\bibinfo  {journal} {Phys. Rev. Lett.}\ }\textbf {\bibinfo {volume} {124}},\ \bibinfo {pages} {020503} (\bibinfo {year} {2020})}\BibitemShut {NoStop}%
\bibitem [{\citenamefont {Pfau}\ \emph {et~al.}(2020)\citenamefont {Pfau}, \citenamefont {Spencer}, \citenamefont {Matthews},\ and\ \citenamefont {Foulkes}}]{pfau2020ferminet}%
  \BibitemOpen
  \bibfield  {author} {\bibinfo {author} {\bibfnamefont {D.}~\bibnamefont {Pfau}}, \bibinfo {author} {\bibfnamefont {J.~S.}\ \bibnamefont {Spencer}}, \bibinfo {author} {\bibfnamefont {A.~G. D.~G.}\ \bibnamefont {Matthews}},\ and\ \bibinfo {author} {\bibfnamefont {W.~M.~C.}\ \bibnamefont {Foulkes}},\ }\bibfield  {title} {\bibinfo {title} {Ab initio solution of the many-electron {S}chr\"odinger equation with deep neural networks},\ }\href {https://doi.org/10.1103/PhysRevResearch.2.033429} {\bibfield  {journal} {\bibinfo  {journal} {Phys. Rev. Res.}\ }\textbf {\bibinfo {volume} {2}},\ \bibinfo {pages} {033429} (\bibinfo {year} {2020})}\BibitemShut {NoStop}%
\bibitem [{\citenamefont {Hermann}\ \emph {et~al.}(2020)\citenamefont {Hermann}, \citenamefont {Schätzle},\ and\ \citenamefont {Noé}}]{hermann2020paulinet}%
  \BibitemOpen
  \bibfield  {author} {\bibinfo {author} {\bibfnamefont {J.}~\bibnamefont {Hermann}}, \bibinfo {author} {\bibfnamefont {Z.}~\bibnamefont {Schätzle}},\ and\ \bibinfo {author} {\bibfnamefont {F.}~\bibnamefont {Noé}},\ }\bibfield  {title} {\bibinfo {title} {Deep-neural-network solution of the electronic {S}chrödinger equation},\ }\href {https://doi.org/10.1038/s41557-020-0544-y} {\bibfield  {journal} {\bibinfo  {journal} {Nature Chemistry}\ }\textbf {\bibinfo {volume} {12}},\ \bibinfo {pages} {891–897} (\bibinfo {year} {2020})}\BibitemShut {NoStop}%
\bibitem [{\citenamefont {Viteritti}\ \emph {et~al.}(2023)\citenamefont {Viteritti}, \citenamefont {Rende},\ and\ \citenamefont {Becca}}]{viteritti2023spinformer}%
  \BibitemOpen
  \bibfield  {author} {\bibinfo {author} {\bibfnamefont {L.~L.}\ \bibnamefont {Viteritti}}, \bibinfo {author} {\bibfnamefont {R.}~\bibnamefont {Rende}},\ and\ \bibinfo {author} {\bibfnamefont {F.}~\bibnamefont {Becca}},\ }\bibfield  {title} {\bibinfo {title} {Transformer variational wave functions for frustrated quantum spin systems},\ }\href {https://doi.org/10.1103/PhysRevLett.130.236401} {\bibfield  {journal} {\bibinfo  {journal} {Phys. Rev. Lett.}\ }\textbf {\bibinfo {volume} {130}},\ \bibinfo {pages} {236401} (\bibinfo {year} {2023})}\BibitemShut {NoStop}%
\bibitem [{\citenamefont {von Glehn}\ \emph {et~al.}(2023)\citenamefont {von Glehn}, \citenamefont {Spencer},\ and\ \citenamefont {Pfau}}]{vonglehn2023psiformer}%
  \BibitemOpen
  \bibfield  {author} {\bibinfo {author} {\bibfnamefont {I.}~\bibnamefont {von Glehn}}, \bibinfo {author} {\bibfnamefont {J.~S.}\ \bibnamefont {Spencer}},\ and\ \bibinfo {author} {\bibfnamefont {D.}~\bibnamefont {Pfau}},\ }\bibfield  {title} {\bibinfo {title} {A self-attention ansatz for ab-initio quantum chemistry},\ }in\ \href {https://openreview.net/forum?id=xveTeHVlF7j} {\emph {\bibinfo {booktitle} {The Eleventh International Conference on Learning Representations}}}\ (\bibinfo {year} {2023})\BibitemShut {NoStop}%
\bibitem [{\citenamefont {Hermann}\ \emph {et~al.}(2023)\citenamefont {Hermann}, \citenamefont {Spencer}, \citenamefont {Choo}, \citenamefont {Mezzacapo}, \citenamefont {Foulkes}, \citenamefont {Pfau}, \citenamefont {Carleo},\ and\ \citenamefont {Noé}}]{hermann2023review}%
  \BibitemOpen
  \bibfield  {author} {\bibinfo {author} {\bibfnamefont {J.}~\bibnamefont {Hermann}}, \bibinfo {author} {\bibfnamefont {J.}~\bibnamefont {Spencer}}, \bibinfo {author} {\bibfnamefont {K.}~\bibnamefont {Choo}}, \bibinfo {author} {\bibfnamefont {A.}~\bibnamefont {Mezzacapo}}, \bibinfo {author} {\bibfnamefont {W.~M.}\ \bibnamefont {Foulkes}}, \bibinfo {author} {\bibfnamefont {D.}~\bibnamefont {Pfau}}, \bibinfo {author} {\bibfnamefont {G.}~\bibnamefont {Carleo}},\ and\ \bibinfo {author} {\bibfnamefont {F.}~\bibnamefont {Noé}},\ }\bibfield  {title} {\bibinfo {title} {Ab initio quantum chemistry with neural-network wavefunctions},\ }\href {https://doi.org/10.1038/s41570-023-00516-8} {\bibfield  {journal} {\bibinfo  {journal} {Nature Reviews Chemistry}\ }\textbf {\bibinfo {volume} {7}},\ \bibinfo {pages} {692–709} (\bibinfo {year} {2023})}\BibitemShut {NoStop}%
\bibitem [{\citenamefont {Lange}\ \emph {et~al.}(2024)\citenamefont {Lange}, \citenamefont {Van~de Walle}, \citenamefont {Abedinnia},\ and\ \citenamefont {Bohrdt}}]{lange2024review}%
  \BibitemOpen
  \bibfield  {author} {\bibinfo {author} {\bibfnamefont {H.}~\bibnamefont {Lange}}, \bibinfo {author} {\bibfnamefont {A.}~\bibnamefont {Van~de Walle}}, \bibinfo {author} {\bibfnamefont {A.}~\bibnamefont {Abedinnia}},\ and\ \bibinfo {author} {\bibfnamefont {A.}~\bibnamefont {Bohrdt}},\ }\bibfield  {title} {\bibinfo {title} {From architectures to applications: a review of neural quantum states},\ }\href {https://doi.org/10.1088/2058-9565/ad7168} {\bibfield  {journal} {\bibinfo  {journal} {Quantum Science and Technology}\ }\textbf {\bibinfo {volume} {9}},\ \bibinfo {pages} {040501} (\bibinfo {year} {2024})}\BibitemShut {NoStop}%
\bibitem [{\citenamefont {Feynman}(1954)}]{feynman1954og}%
  \BibitemOpen
  \bibfield  {author} {\bibinfo {author} {\bibfnamefont {R.~P.}\ \bibnamefont {Feynman}},\ }\bibfield  {title} {\bibinfo {title} {Atomic theory of the two-fluid model of liquid helium},\ }\href {https://doi.org/10.1103/PhysRev.94.262} {\bibfield  {journal} {\bibinfo  {journal} {Phys. Rev.}\ }\textbf {\bibinfo {volume} {94}},\ \bibinfo {pages} {262} (\bibinfo {year} {1954})}\BibitemShut {NoStop}%
\bibitem [{\citenamefont {McMillan}(1965)}]{mcmillan1965og}%
  \BibitemOpen
  \bibfield  {author} {\bibinfo {author} {\bibfnamefont {W.~L.}\ \bibnamefont {McMillan}},\ }\bibfield  {title} {\bibinfo {title} {Ground state of liquid {${\mathrm{He}}^{4}$}},\ }\href {https://doi.org/10.1103/PhysRev.138.A442} {\bibfield  {journal} {\bibinfo  {journal} {Phys. Rev.}\ }\textbf {\bibinfo {volume} {138}},\ \bibinfo {pages} {A442} (\bibinfo {year} {1965})}\BibitemShut {NoStop}%
\bibitem [{\citenamefont {Ceperley}\ \emph {et~al.}(1977)\citenamefont {Ceperley}, \citenamefont {Chester},\ and\ \citenamefont {Kalos}}]{ceperley1977first}%
  \BibitemOpen
  \bibfield  {author} {\bibinfo {author} {\bibfnamefont {D.}~\bibnamefont {Ceperley}}, \bibinfo {author} {\bibfnamefont {G.~V.}\ \bibnamefont {Chester}},\ and\ \bibinfo {author} {\bibfnamefont {M.~H.}\ \bibnamefont {Kalos}},\ }\bibfield  {title} {\bibinfo {title} {{M}onte {C}arlo simulation of a many-fermion study},\ }\href {https://doi.org/10.1103/PhysRevB.16.3081} {\bibfield  {journal} {\bibinfo  {journal} {Phys. Rev. B}\ }\textbf {\bibinfo {volume} {16}},\ \bibinfo {pages} {3081} (\bibinfo {year} {1977})}\BibitemShut {NoStop}%
\bibitem [{\citenamefont {Ceperley}\ and\ \citenamefont {Alder}(1980)}]{ceperley1980ground}%
  \BibitemOpen
  \bibfield  {author} {\bibinfo {author} {\bibfnamefont {D.~M.}\ \bibnamefont {Ceperley}}\ and\ \bibinfo {author} {\bibfnamefont {B.~J.}\ \bibnamefont {Alder}},\ }\bibfield  {title} {\bibinfo {title} {Ground state of the electron gas by a stochastic method},\ }\href {https://doi.org/10.1103/PhysRevLett.45.566} {\bibfield  {journal} {\bibinfo  {journal} {Phys. Rev. Lett.}\ }\textbf {\bibinfo {volume} {45}},\ \bibinfo {pages} {566} (\bibinfo {year} {1980})}\BibitemShut {NoStop}%
\bibitem [{\citenamefont {Tanatar}\ and\ \citenamefont {Ceperley}(1989)}]{tanatar1989twodeg}%
  \BibitemOpen
  \bibfield  {author} {\bibinfo {author} {\bibfnamefont {B.}~\bibnamefont {Tanatar}}\ and\ \bibinfo {author} {\bibfnamefont {D.~M.}\ \bibnamefont {Ceperley}},\ }\bibfield  {title} {\bibinfo {title} {Ground state of the two-dimensional electron gas},\ }\href {https://doi.org/10.1103/PhysRevB.39.5005} {\bibfield  {journal} {\bibinfo  {journal} {Phys. Rev. B}\ }\textbf {\bibinfo {volume} {39}},\ \bibinfo {pages} {5005} (\bibinfo {year} {1989})}\BibitemShut {NoStop}%
\bibitem [{\citenamefont {Foulkes}\ \emph {et~al.}(2001)\citenamefont {Foulkes}, \citenamefont {Mitas}, \citenamefont {Needs},\ and\ \citenamefont {Rajagopal}}]{foulkes2001review}%
  \BibitemOpen
  \bibfield  {author} {\bibinfo {author} {\bibfnamefont {W.~M.~C.}\ \bibnamefont {Foulkes}}, \bibinfo {author} {\bibfnamefont {L.}~\bibnamefont {Mitas}}, \bibinfo {author} {\bibfnamefont {R.~J.}\ \bibnamefont {Needs}},\ and\ \bibinfo {author} {\bibfnamefont {G.}~\bibnamefont {Rajagopal}},\ }\bibfield  {title} {\bibinfo {title} {Quantum {M}onte {C}arlo simulations of solids},\ }\href {https://doi.org/10.1103/RevModPhys.73.33} {\bibfield  {journal} {\bibinfo  {journal} {Rev. Mod. Phys.}\ }\textbf {\bibinfo {volume} {73}},\ \bibinfo {pages} {33} (\bibinfo {year} {2001})}\BibitemShut {NoStop}%
\bibitem [{\citenamefont {Needs}\ \emph {et~al.}(2009)\citenamefont {Needs}, \citenamefont {Towler}, \citenamefont {Drummond},\ and\ \citenamefont {López~Ríos}}]{needs2010review}%
  \BibitemOpen
  \bibfield  {author} {\bibinfo {author} {\bibfnamefont {R.~J.}\ \bibnamefont {Needs}}, \bibinfo {author} {\bibfnamefont {M.~D.}\ \bibnamefont {Towler}}, \bibinfo {author} {\bibfnamefont {N.~D.}\ \bibnamefont {Drummond}},\ and\ \bibinfo {author} {\bibfnamefont {P.}~\bibnamefont {López~Ríos}},\ }\bibfield  {title} {\bibinfo {title} {Continuum variational and diffusion quantum {M}onte {C}arlo calculations},\ }\href {https://doi.org/10.1088/0953-8984/22/2/023201} {\bibfield  {journal} {\bibinfo  {journal} {Journal of Physics: Condensed Matter}\ }\textbf {\bibinfo {volume} {22}},\ \bibinfo {pages} {023201} (\bibinfo {year} {2009})}\BibitemShut {NoStop}%
\bibitem [{\citenamefont {Spencer}\ \emph {et~al.}(2020)\citenamefont {Spencer}, \citenamefont {Pfau}, \citenamefont {Botev},\ and\ \citenamefont {Foulkes}}]{spencer2020better}%
  \BibitemOpen
  \bibfield  {author} {\bibinfo {author} {\bibfnamefont {J.~S.}\ \bibnamefont {Spencer}}, \bibinfo {author} {\bibfnamefont {D.}~\bibnamefont {Pfau}}, \bibinfo {author} {\bibfnamefont {A.}~\bibnamefont {Botev}},\ and\ \bibinfo {author} {\bibfnamefont {W.~M.~C.}\ \bibnamefont {Foulkes}},\ }\href {https://arxiv.org/abs/2011.07125} {\bibinfo {title} {Better, faster fermionic neural networks}} (\bibinfo {year} {2020}),\ \Eprint {https://arxiv.org/abs/2011.07125} {arXiv:2011.07125 [physics.comp-ph]} \BibitemShut {NoStop}%
\bibitem [{\citenamefont {Moss}\ \emph {et~al.}(2025)\citenamefont {Moss}, \citenamefont {Orfi}, \citenamefont {Roth}, \citenamefont {Sengupta}, \citenamefont {Georges}, \citenamefont {Sels}, \citenamefont {Dawid},\ and\ \citenamefont {Valenti}}]{moss2025double}%
  \BibitemOpen
  \bibfield  {author} {\bibinfo {author} {\bibfnamefont {M.~S.}\ \bibnamefont {Moss}}, \bibinfo {author} {\bibfnamefont {A.}~\bibnamefont {Orfi}}, \bibinfo {author} {\bibfnamefont {C.}~\bibnamefont {Roth}}, \bibinfo {author} {\bibfnamefont {A.~M.}\ \bibnamefont {Sengupta}}, \bibinfo {author} {\bibfnamefont {A.}~\bibnamefont {Georges}}, \bibinfo {author} {\bibfnamefont {D.}~\bibnamefont {Sels}}, \bibinfo {author} {\bibfnamefont {A.}~\bibnamefont {Dawid}},\ and\ \bibinfo {author} {\bibfnamefont {A.}~\bibnamefont {Valenti}},\ }\href {https://arxiv.org/abs/2508.00068} {\bibinfo {title} {Double descent: When do neural quantum states generalize?}} (\bibinfo {year} {2025}),\ \Eprint {https://arxiv.org/abs/2508.00068} {arXiv:2508.00068 [cond-mat.dis-nn]} \BibitemShut {NoStop}%
\bibitem [{\citenamefont {Sutskever}\ \emph {et~al.}(2013)\citenamefont {Sutskever}, \citenamefont {Martens}, \citenamefont {Dahl},\ and\ \citenamefont {Hinton}}]{sutskever2013importance}%
  \BibitemOpen
  \bibfield  {author} {\bibinfo {author} {\bibfnamefont {I.}~\bibnamefont {Sutskever}}, \bibinfo {author} {\bibfnamefont {J.}~\bibnamefont {Martens}}, \bibinfo {author} {\bibfnamefont {G.}~\bibnamefont {Dahl}},\ and\ \bibinfo {author} {\bibfnamefont {G.}~\bibnamefont {Hinton}},\ }\bibfield  {title} {\bibinfo {title} {On the importance of initialization and momentum in deep learning},\ }in\ \href {https://proceedings.mlr.press/v28/sutskever13.html} {\emph {\bibinfo {booktitle} {Proceedings of the 30th International Conference on Machine Learning}}},\ \bibinfo {series} {Proceedings of Machine Learning Research}, Vol.~\bibinfo {volume} {28},\ \bibinfo {editor} {edited by\ \bibinfo {editor} {\bibfnamefont {S.}~\bibnamefont {Dasgupta}}\ and\ \bibinfo {editor} {\bibfnamefont {D.}~\bibnamefont {McAllester}}}\ (\bibinfo  {publisher} {PMLR},\ \bibinfo {address} {Atlanta, Georgia, USA},\ \bibinfo {year} {2013})\ pp.\ \bibinfo {pages} {1139--1147}\BibitemShut {NoStop}%
\bibitem [{\citenamefont {Kingma}\ and\ \citenamefont {Ba}(2015)}]{kingma2015adam}%
  \BibitemOpen
  \bibfield  {author} {\bibinfo {author} {\bibfnamefont {D.~P.}\ \bibnamefont {Kingma}}\ and\ \bibinfo {author} {\bibfnamefont {J.}~\bibnamefont {Ba}},\ }\bibfield  {title} {\bibinfo {title} {{A}dam: A method for stochastic optimization},\ }in\ \href {https://arxiv.org/abs/1412.6980} {\emph {\bibinfo {booktitle} {International Conference on Learning Representations}}}\ (\bibinfo {year} {2015})\BibitemShut {NoStop}%
\bibitem [{\citenamefont {LeCun}\ \emph {et~al.}(2015)\citenamefont {LeCun}, \citenamefont {Bengio},\ and\ \citenamefont {Hinton}}]{lecun2015deep}%
  \BibitemOpen
  \bibfield  {author} {\bibinfo {author} {\bibfnamefont {Y.}~\bibnamefont {LeCun}}, \bibinfo {author} {\bibfnamefont {Y.}~\bibnamefont {Bengio}},\ and\ \bibinfo {author} {\bibfnamefont {G.}~\bibnamefont {Hinton}},\ }\bibfield  {title} {\bibinfo {title} {Deep learning},\ }\href {https://doi.org/10.1038/nature14539} {\bibfield  {journal} {\bibinfo  {journal} {Nature}\ }\textbf {\bibinfo {volume} {521}},\ \bibinfo {pages} {436} (\bibinfo {year} {2015})}\BibitemShut {NoStop}%
\bibitem [{\citenamefont {Amari}(1998)}]{amari1998natural}%
  \BibitemOpen
  \bibfield  {author} {\bibinfo {author} {\bibfnamefont {S.-i.}\ \bibnamefont {Amari}},\ }\bibfield  {title} {\bibinfo {title} {Natural gradient works efficiently in learning},\ }\href {https://doi.org/10.1162/089976698300017746} {\bibfield  {journal} {\bibinfo  {journal} {Neural Computation}\ }\textbf {\bibinfo {volume} {10}},\ \bibinfo {pages} {251} (\bibinfo {year} {1998})},\ \Eprint {https://arxiv.org/abs/https://direct.mit.edu/neco/article-pdf/10/2/251/813415/089976698300017746.pdf} {https://direct.mit.edu/neco/article-pdf/10/2/251/813415/089976698300017746.pdf} \BibitemShut {NoStop}%
\bibitem [{\citenamefont {Sorella}(1998)}]{sorella1998green}%
  \BibitemOpen
  \bibfield  {author} {\bibinfo {author} {\bibfnamefont {S.}~\bibnamefont {Sorella}},\ }\bibfield  {title} {\bibinfo {title} {Green function {M}onte {C}arlo with stochastic reconfiguration},\ }\href {https://doi.org/10.1103/PhysRevLett.80.4558} {\bibfield  {journal} {\bibinfo  {journal} {Phys. Rev. Lett.}\ }\textbf {\bibinfo {volume} {80}},\ \bibinfo {pages} {4558} (\bibinfo {year} {1998})}\BibitemShut {NoStop}%
\bibitem [{\citenamefont {Sorella}(2001)}]{sorella2001lanczos}%
  \BibitemOpen
  \bibfield  {author} {\bibinfo {author} {\bibfnamefont {S.}~\bibnamefont {Sorella}},\ }\bibfield  {title} {\bibinfo {title} {Generalized {L}anczos algorithm for variational quantum {M}onte {C}arlo},\ }\href {https://doi.org/10.1103/PhysRevB.64.024512} {\bibfield  {journal} {\bibinfo  {journal} {Phys. Rev. B}\ }\textbf {\bibinfo {volume} {64}},\ \bibinfo {pages} {024512} (\bibinfo {year} {2001})}\BibitemShut {NoStop}%
\bibitem [{\citenamefont {Martens}\ and\ \citenamefont {Grosse}(2015)}]{martens2015kfac}%
  \BibitemOpen
  \bibfield  {author} {\bibinfo {author} {\bibfnamefont {J.}~\bibnamefont {Martens}}\ and\ \bibinfo {author} {\bibfnamefont {R.}~\bibnamefont {Grosse}},\ }\bibfield  {title} {\bibinfo {title} {Optimizing neural networks with {K}ronecker-factored approximate curvature},\ }in\ \href {https://proceedings.mlr.press/v37/martens15.html} {\emph {\bibinfo {booktitle} {Proceedings of the 32nd International Conference on Machine Learning}}},\ \bibinfo {series} {Proceedings of Machine Learning Research}, Vol.~\bibinfo {volume} {37},\ \bibinfo {editor} {edited by\ \bibinfo {editor} {\bibfnamefont {F.}~\bibnamefont {Bach}}\ and\ \bibinfo {editor} {\bibfnamefont {D.}~\bibnamefont {Blei}}}\ (\bibinfo  {publisher} {PMLR},\ \bibinfo {address} {Lille, France},\ \bibinfo {year} {2015})\ pp.\ \bibinfo {pages} {2408--2417}\BibitemShut {NoStop}%
\bibitem [{\citenamefont {Stokes}\ \emph {et~al.}(2020)\citenamefont {Stokes}, \citenamefont {Izaac}, \citenamefont {Killoran},\ and\ \citenamefont {Carleo}}]{stokes2020natural}%
  \BibitemOpen
  \bibfield  {author} {\bibinfo {author} {\bibfnamefont {J.}~\bibnamefont {Stokes}}, \bibinfo {author} {\bibfnamefont {J.}~\bibnamefont {Izaac}}, \bibinfo {author} {\bibfnamefont {N.}~\bibnamefont {Killoran}},\ and\ \bibinfo {author} {\bibfnamefont {G.}~\bibnamefont {Carleo}},\ }\bibfield  {title} {\bibinfo {title} {Quantum {N}atural {G}radient},\ }\href {https://doi.org/10.22331/q-2020-05-25-269} {\bibfield  {journal} {\bibinfo  {journal} {{Quantum}}\ }\textbf {\bibinfo {volume} {4}},\ \bibinfo {pages} {269} (\bibinfo {year} {2020})}\BibitemShut {NoStop}%
\bibitem [{\citenamefont {Lange}\ \emph {et~al.}(2025)\citenamefont {Lange}, \citenamefont {Bornet}, \citenamefont {Emperauger}, \citenamefont {Chen}, \citenamefont {Lahaye}, \citenamefont {Kienle}, \citenamefont {Browaeys},\ and\ \citenamefont {Bohrdt}}]{lange2025hybrid}%
  \BibitemOpen
  \bibfield  {author} {\bibinfo {author} {\bibfnamefont {H.}~\bibnamefont {Lange}}, \bibinfo {author} {\bibfnamefont {G.}~\bibnamefont {Bornet}}, \bibinfo {author} {\bibfnamefont {G.}~\bibnamefont {Emperauger}}, \bibinfo {author} {\bibfnamefont {C.}~\bibnamefont {Chen}}, \bibinfo {author} {\bibfnamefont {T.}~\bibnamefont {Lahaye}}, \bibinfo {author} {\bibfnamefont {S.}~\bibnamefont {Kienle}}, \bibinfo {author} {\bibfnamefont {A.}~\bibnamefont {Browaeys}},\ and\ \bibinfo {author} {\bibfnamefont {A.}~\bibnamefont {Bohrdt}},\ }\bibfield  {title} {\bibinfo {title} {Transformer neural networks and quantum simulators: a hybrid approach for simulating strongly correlated systems},\ }\href {https://doi.org/10.22331/q-2025-03-26-1675} {\bibfield  {journal} {\bibinfo  {journal} {{Quantum}}\ }\textbf {\bibinfo {volume} {9}},\ \bibinfo {pages} {1675} (\bibinfo {year} {2025})}\BibitemShut {NoStop}%
\bibitem [{\citenamefont {Garipov}\ \emph {et~al.}(2018)\citenamefont {Garipov}, \citenamefont {Izmailov}, \citenamefont {Podoprikhin}, \citenamefont {Vetrov},\ and\ \citenamefont {Wilson}}]{garipov2018loss}%
  \BibitemOpen
  \bibfield  {author} {\bibinfo {author} {\bibfnamefont {T.}~\bibnamefont {Garipov}}, \bibinfo {author} {\bibfnamefont {P.}~\bibnamefont {Izmailov}}, \bibinfo {author} {\bibfnamefont {D.}~\bibnamefont {Podoprikhin}}, \bibinfo {author} {\bibfnamefont {D.~P.}\ \bibnamefont {Vetrov}},\ and\ \bibinfo {author} {\bibfnamefont {A.~G.}\ \bibnamefont {Wilson}},\ }\bibfield  {title} {\bibinfo {title} {Loss surfaces, mode connectivity, and fast ensembling of {DNNs}},\ }in\ \href {https://proceedings.neurips.cc/paper_files/paper/2018/file/be3087e74e9100d4bc4c6268cdbe8456-Paper.pdf} {\emph {\bibinfo {booktitle} {Advances in Neural Information Processing Systems}}},\ Vol.~\bibinfo {volume} {31},\ \bibinfo {editor} {edited by\ \bibinfo {editor} {\bibfnamefont {S.}~\bibnamefont {Bengio}}, \bibinfo {editor} {\bibfnamefont {H.}~\bibnamefont {Wallach}}, \bibinfo {editor} {\bibfnamefont {H.}~\bibnamefont {Larochelle}}, \bibinfo {editor} {\bibfnamefont {K.}~\bibnamefont {Grauman}}, \bibinfo {editor} {\bibfnamefont
  {N.}~\bibnamefont {Cesa-Bianchi}},\ and\ \bibinfo {editor} {\bibfnamefont {R.}~\bibnamefont {Garnett}}}\ (\bibinfo  {publisher} {Curran Associates, Inc.},\ \bibinfo {year} {2018})\BibitemShut {NoStop}%
\bibitem [{\citenamefont {Draxler}\ \emph {et~al.}(2018)\citenamefont {Draxler}, \citenamefont {Veschgini}, \citenamefont {Salmhofer},\ and\ \citenamefont {Hamprecht}}]{draxler2018essentially}%
  \BibitemOpen
  \bibfield  {author} {\bibinfo {author} {\bibfnamefont {F.}~\bibnamefont {Draxler}}, \bibinfo {author} {\bibfnamefont {K.}~\bibnamefont {Veschgini}}, \bibinfo {author} {\bibfnamefont {M.}~\bibnamefont {Salmhofer}},\ and\ \bibinfo {author} {\bibfnamefont {F.}~\bibnamefont {Hamprecht}},\ }\bibfield  {title} {\bibinfo {title} {Essentially no barriers in neural network energy landscape},\ }in\ \href {https://proceedings.mlr.press/v80/draxler18a.html} {\emph {\bibinfo {booktitle} {Proceedings of the 35th International Conference on Machine Learning}}},\ \bibinfo {series} {Proceedings of Machine Learning Research}, Vol.~\bibinfo {volume} {80},\ \bibinfo {editor} {edited by\ \bibinfo {editor} {\bibfnamefont {J.}~\bibnamefont {Dy}}\ and\ \bibinfo {editor} {\bibfnamefont {A.}~\bibnamefont {Krause}}}\ (\bibinfo  {publisher} {PMLR},\ \bibinfo {year} {2018})\ pp.\ \bibinfo {pages} {1309--1318}\BibitemShut {NoStop}%
\bibitem [{\citenamefont {J{\'o}nsson}\ \emph {et~al.}(1998)\citenamefont {J{\'o}nsson}, \citenamefont {Mills},\ and\ \citenamefont {Jacobsen}}]{jonsson1998neb}%
  \BibitemOpen
  \bibfield  {author} {\bibinfo {author} {\bibfnamefont {H.}~\bibnamefont {J{\'o}nsson}}, \bibinfo {author} {\bibfnamefont {G.}~\bibnamefont {Mills}},\ and\ \bibinfo {author} {\bibfnamefont {K.~W.}\ \bibnamefont {Jacobsen}},\ }\bibfield  {title} {\bibinfo {title} {Nudged elastic band method for finding minimum energy paths of transitions},\ }in\ \href {https://doi.org/10.1142/9789812839664_0016} {\emph {\bibinfo {booktitle} {Classical and Quantum Dynamics in Condensed Phase Simulations}}},\ \bibinfo {editor} {edited by\ \bibinfo {editor} {\bibfnamefont {B.~J.}\ \bibnamefont {Berne}}, \bibinfo {editor} {\bibfnamefont {G.}~\bibnamefont {Ciccotti}},\ and\ \bibinfo {editor} {\bibfnamefont {D.~F.}\ \bibnamefont {Coker}}}\ (\bibinfo  {publisher} {World Scientific},\ \bibinfo {address} {Singapore},\ \bibinfo {year} {1998})\ p.\ \bibinfo {pages} {385–404}\BibitemShut {NoStop}%
\bibitem [{\citenamefont {Kolsbjerg}\ \emph {et~al.}(2016)\citenamefont {Kolsbjerg}, \citenamefont {Groves},\ and\ \citenamefont {Hammer}}]{kolsbjerg2016autoneb}%
  \BibitemOpen
  \bibfield  {author} {\bibinfo {author} {\bibfnamefont {E.~L.}\ \bibnamefont {Kolsbjerg}}, \bibinfo {author} {\bibfnamefont {M.~N.}\ \bibnamefont {Groves}},\ and\ \bibinfo {author} {\bibfnamefont {B.}~\bibnamefont {Hammer}},\ }\bibfield  {title} {\bibinfo {title} {An automated nudged elastic band method},\ }\href {https://doi.org/10.1063/1.4961868} {\bibfield  {journal} {\bibinfo  {journal} {The Journal of Chemical Physics}\ }\textbf {\bibinfo {volume} {145}},\ \bibinfo {pages} {094107} (\bibinfo {year} {2016})}\BibitemShut {NoStop}%
\bibitem [{\citenamefont {Chen}\ and\ \citenamefont {Heyl}(2024)}]{chen2024empowering}%
  \BibitemOpen
  \bibfield  {author} {\bibinfo {author} {\bibfnamefont {A.}~\bibnamefont {Chen}}\ and\ \bibinfo {author} {\bibfnamefont {M.}~\bibnamefont {Heyl}},\ }\bibfield  {title} {\bibinfo {title} {Empowering deep neural quantum states through efficient optimization},\ }\href {https://doi.org/10.1038/s41567-024-02566-1} {\bibfield  {journal} {\bibinfo  {journal} {Nature Physics}\ }\textbf {\bibinfo {volume} {20}},\ \bibinfo {pages} {1476} (\bibinfo {year} {2024})}\BibitemShut {NoStop}%
\bibitem [{\citenamefont {Rende}\ \emph {et~al.}(2024)\citenamefont {Rende}, \citenamefont {Viteritti}, \citenamefont {Bardone}, \citenamefont {Becca},\ and\ \citenamefont {Goldt}}]{rende2024simple}%
  \BibitemOpen
  \bibfield  {author} {\bibinfo {author} {\bibfnamefont {R.}~\bibnamefont {Rende}}, \bibinfo {author} {\bibfnamefont {L.~L.}\ \bibnamefont {Viteritti}}, \bibinfo {author} {\bibfnamefont {L.}~\bibnamefont {Bardone}}, \bibinfo {author} {\bibfnamefont {F.}~\bibnamefont {Becca}},\ and\ \bibinfo {author} {\bibfnamefont {S.}~\bibnamefont {Goldt}},\ }\bibfield  {title} {\bibinfo {title} {A simple linear algebra identity to optimize large-scale neural network quantum states},\ }\href {https://doi.org/10.1038/s42005-024-01732-4} {\bibfield  {journal} {\bibinfo  {journal} {Communications Physics}\ }\textbf {\bibinfo {volume} {7}},\ \bibinfo {pages} {260} (\bibinfo {year} {2024})}\BibitemShut {NoStop}%
\bibitem [{\citenamefont {Vaswani}\ \emph {et~al.}(2017)\citenamefont {Vaswani}, \citenamefont {Shazeer}, \citenamefont {Parmar}, \citenamefont {Uszkoreit}, \citenamefont {Jones}, \citenamefont {Gomez}, \citenamefont {Kaiser},\ and\ \citenamefont {Polosukhin}}]{vaswani2017attention}%
  \BibitemOpen
  \bibfield  {author} {\bibinfo {author} {\bibfnamefont {A.}~\bibnamefont {Vaswani}}, \bibinfo {author} {\bibfnamefont {N.}~\bibnamefont {Shazeer}}, \bibinfo {author} {\bibfnamefont {N.}~\bibnamefont {Parmar}}, \bibinfo {author} {\bibfnamefont {J.}~\bibnamefont {Uszkoreit}}, \bibinfo {author} {\bibfnamefont {L.}~\bibnamefont {Jones}}, \bibinfo {author} {\bibfnamefont {A.~N.}\ \bibnamefont {Gomez}}, \bibinfo {author} {\bibfnamefont {L.~u.}\ \bibnamefont {Kaiser}},\ and\ \bibinfo {author} {\bibfnamefont {I.}~\bibnamefont {Polosukhin}},\ }\bibfield  {title} {\bibinfo {title} {Attention is all you need},\ }in\ \href {https://proceedings.neurips.cc/paper_files/paper/2017/file/3f5ee243547dee91fbd053c1c4a845aa-Paper.pdf} {\emph {\bibinfo {booktitle} {Advances in Neural Information Processing Systems}}},\ Vol.~\bibinfo {volume} {30},\ \bibinfo {editor} {edited by\ \bibinfo {editor} {\bibfnamefont {I.}~\bibnamefont {Guyon}}, \bibinfo {editor} {\bibfnamefont {U.~V.}\ \bibnamefont {Luxburg}}, \bibinfo {editor}
  {\bibfnamefont {S.}~\bibnamefont {Bengio}}, \bibinfo {editor} {\bibfnamefont {H.}~\bibnamefont {Wallach}}, \bibinfo {editor} {\bibfnamefont {R.}~\bibnamefont {Fergus}}, \bibinfo {editor} {\bibfnamefont {S.}~\bibnamefont {Vishwanathan}},\ and\ \bibinfo {editor} {\bibfnamefont {R.}~\bibnamefont {Garnett}}}\ (\bibinfo  {publisher} {Curran Associates, Inc.},\ \bibinfo {year} {2017})\BibitemShut {NoStop}%
\bibitem [{\citenamefont {Dosovitskiy}\ \emph {et~al.}(2021)\citenamefont {Dosovitskiy}, \citenamefont {Beyer}, \citenamefont {Kolesnikov}, \citenamefont {Weissenborn}, \citenamefont {Zhai}, \citenamefont {Unterthiner}, \citenamefont {Dehghani}, \citenamefont {Minderer}, \citenamefont {Heigold}, \citenamefont {Gelly}, \citenamefont {Uszkoreit},\ and\ \citenamefont {Houlsby}}]{dosovitskiy2021vit}%
  \BibitemOpen
  \bibfield  {author} {\bibinfo {author} {\bibfnamefont {A.}~\bibnamefont {Dosovitskiy}}, \bibinfo {author} {\bibfnamefont {L.}~\bibnamefont {Beyer}}, \bibinfo {author} {\bibfnamefont {A.}~\bibnamefont {Kolesnikov}}, \bibinfo {author} {\bibfnamefont {D.}~\bibnamefont {Weissenborn}}, \bibinfo {author} {\bibfnamefont {X.}~\bibnamefont {Zhai}}, \bibinfo {author} {\bibfnamefont {T.}~\bibnamefont {Unterthiner}}, \bibinfo {author} {\bibfnamefont {M.}~\bibnamefont {Dehghani}}, \bibinfo {author} {\bibfnamefont {M.}~\bibnamefont {Minderer}}, \bibinfo {author} {\bibfnamefont {G.}~\bibnamefont {Heigold}}, \bibinfo {author} {\bibfnamefont {S.}~\bibnamefont {Gelly}}, \bibinfo {author} {\bibfnamefont {J.}~\bibnamefont {Uszkoreit}},\ and\ \bibinfo {author} {\bibfnamefont {N.}~\bibnamefont {Houlsby}},\ }\bibfield  {title} {\bibinfo {title} {An image is worth 16x16 words: {T}ransformers for image recognition at scale},\ }in\ \href {https://openreview.net/forum?id=YicbFdNTTy} {\emph {\bibinfo {booktitle} {International
  Conference on Learning Representations}}}\ (\bibinfo {year} {2021})\BibitemShut {NoStop}%
\bibitem [{\citenamefont {Xiong}\ \emph {et~al.}(2020)\citenamefont {Xiong}, \citenamefont {Yang}, \citenamefont {He}, \citenamefont {Zheng}, \citenamefont {Zheng}, \citenamefont {Xing}, \citenamefont {Zhang}, \citenamefont {Lan}, \citenamefont {Wang},\ and\ \citenamefont {Liu}}]{xiong2020prenorm}%
  \BibitemOpen
  \bibfield  {author} {\bibinfo {author} {\bibfnamefont {R.}~\bibnamefont {Xiong}}, \bibinfo {author} {\bibfnamefont {Y.}~\bibnamefont {Yang}}, \bibinfo {author} {\bibfnamefont {D.}~\bibnamefont {He}}, \bibinfo {author} {\bibfnamefont {K.}~\bibnamefont {Zheng}}, \bibinfo {author} {\bibfnamefont {S.}~\bibnamefont {Zheng}}, \bibinfo {author} {\bibfnamefont {C.}~\bibnamefont {Xing}}, \bibinfo {author} {\bibfnamefont {H.}~\bibnamefont {Zhang}}, \bibinfo {author} {\bibfnamefont {Y.}~\bibnamefont {Lan}}, \bibinfo {author} {\bibfnamefont {L.}~\bibnamefont {Wang}},\ and\ \bibinfo {author} {\bibfnamefont {T.}~\bibnamefont {Liu}},\ }\bibfield  {title} {\bibinfo {title} {On layer normalization in the {T}ransformer architecture},\ }in\ \href {https://proceedings.mlr.press/v119/xiong20b.html} {\emph {\bibinfo {booktitle} {Proceedings of the 37th International Conference on Machine Learning}}},\ \bibinfo {series} {Proceedings of Machine Learning Research}, Vol.\ \bibinfo {volume} {119},\ \bibinfo {editor} {edited by\
  \bibinfo {editor} {\bibfnamefont {H.~D.}\ \bibnamefont {III}}\ and\ \bibinfo {editor} {\bibfnamefont {A.}~\bibnamefont {Singh}}}\ (\bibinfo  {publisher} {PMLR},\ \bibinfo {year} {2020})\ pp.\ \bibinfo {pages} {10524--10533}\BibitemShut {NoStop}%
\bibitem [{\citenamefont {He}\ \emph {et~al.}(2016)\citenamefont {He}, \citenamefont {Zhang}, \citenamefont {Ren},\ and\ \citenamefont {Sun}}]{he2016prenorm}%
  \BibitemOpen
  \bibfield  {author} {\bibinfo {author} {\bibfnamefont {K.}~\bibnamefont {He}}, \bibinfo {author} {\bibfnamefont {X.}~\bibnamefont {Zhang}}, \bibinfo {author} {\bibfnamefont {S.}~\bibnamefont {Ren}},\ and\ \bibinfo {author} {\bibfnamefont {J.}~\bibnamefont {Sun}},\ }\bibfield  {title} {\bibinfo {title} {Identity mappings in deep residual networks},\ }in\ \href@noop {} {\emph {\bibinfo {booktitle} {Computer Vision -- ECCV 2016}}},\ \bibinfo {editor} {edited by\ \bibinfo {editor} {\bibfnamefont {B.}~\bibnamefont {Leibe}}, \bibinfo {editor} {\bibfnamefont {J.}~\bibnamefont {Matas}}, \bibinfo {editor} {\bibfnamefont {N.}~\bibnamefont {Sebe}},\ and\ \bibinfo {editor} {\bibfnamefont {M.}~\bibnamefont {Welling}}}\ (\bibinfo  {publisher} {Springer International Publishing},\ \bibinfo {address} {Cham},\ \bibinfo {year} {2016})\ pp.\ \bibinfo {pages} {630--645}\BibitemShut {NoStop}%
\bibitem [{Note1()}]{Note1}%
  \BibitemOpen
  \bibinfo {note} {We specialized to real wavefunctions here, but everything can be easily generalized to complex wavefunctions following~\cite {chen2024empowering,rende2024simple}.}\BibitemShut {Stop}%
\bibitem [{\citenamefont {Fu}\ \emph {et~al.}(2024)\citenamefont {Fu}, \citenamefont {Ren},\ and\ \citenamefont {Chen}}]{fu2024variance}%
  \BibitemOpen
  \bibfield  {author} {\bibinfo {author} {\bibfnamefont {W.}~\bibnamefont {Fu}}, \bibinfo {author} {\bibfnamefont {W.}~\bibnamefont {Ren}},\ and\ \bibinfo {author} {\bibfnamefont {J.}~\bibnamefont {Chen}},\ }\bibfield  {title} {\bibinfo {title} {Variance extrapolation method for neural-network variational {M}onte {C}arlo},\ }\href {https://doi.org/10.1088/2632-2153/ad1f75} {\bibfield  {journal} {\bibinfo  {journal} {Machine Learning: Science and Technology}\ }\textbf {\bibinfo {volume} {5}},\ \bibinfo {pages} {015016} (\bibinfo {year} {2024})}\BibitemShut {NoStop}%
\bibitem [{\citenamefont {Kuditipudi}\ \emph {et~al.}(2019)\citenamefont {Kuditipudi}, \citenamefont {Wang}, \citenamefont {Lee}, \citenamefont {Zhang}, \citenamefont {Li}, \citenamefont {Hu}, \citenamefont {Ge},\ and\ \citenamefont {Arora}}]{kuditipudi2019noise}%
  \BibitemOpen
  \bibfield  {author} {\bibinfo {author} {\bibfnamefont {R.}~\bibnamefont {Kuditipudi}}, \bibinfo {author} {\bibfnamefont {X.}~\bibnamefont {Wang}}, \bibinfo {author} {\bibfnamefont {H.}~\bibnamefont {Lee}}, \bibinfo {author} {\bibfnamefont {Y.}~\bibnamefont {Zhang}}, \bibinfo {author} {\bibfnamefont {Z.}~\bibnamefont {Li}}, \bibinfo {author} {\bibfnamefont {W.}~\bibnamefont {Hu}}, \bibinfo {author} {\bibfnamefont {R.}~\bibnamefont {Ge}},\ and\ \bibinfo {author} {\bibfnamefont {S.}~\bibnamefont {Arora}},\ }\bibfield  {title} {\bibinfo {title} {Explaining landscape connectivity of low-cost solutions for multilayer nets},\ }in\ \href {https://proceedings.neurips.cc/paper_files/paper/2019/file/46a4378f835dc8040c8057beb6a2da52-Paper.pdf} {\emph {\bibinfo {booktitle} {Advances in Neural Information Processing Systems}}},\ Vol.~\bibinfo {volume} {32},\ \bibinfo {editor} {edited by\ \bibinfo {editor} {\bibfnamefont {H.}~\bibnamefont {Wallach}}, \bibinfo {editor} {\bibfnamefont {H.}~\bibnamefont {Larochelle}},
  \bibinfo {editor} {\bibfnamefont {A.}~\bibnamefont {Beygelzimer}}, \bibinfo {editor} {\bibfnamefont {F.}~\bibnamefont {d\textquotesingle Alch\'{e}-Buc}}, \bibinfo {editor} {\bibfnamefont {E.}~\bibnamefont {Fox}},\ and\ \bibinfo {editor} {\bibfnamefont {R.}~\bibnamefont {Garnett}}}\ (\bibinfo  {publisher} {Curran Associates, Inc.},\ \bibinfo {year} {2019})\BibitemShut {NoStop}%
\bibitem [{\citenamefont {Hernandes}\ \emph {et~al.}(2025)\citenamefont {Hernandes}, \citenamefont {Spriggs},\ and\ \citenamefont {Greplova}}]{hernandes2025mode}%
  \BibitemOpen
  \bibfield  {author} {\bibinfo {author} {\bibfnamefont {V.}~\bibnamefont {Hernandes}}, \bibinfo {author} {\bibfnamefont {T.}~\bibnamefont {Spriggs}},\ and\ \bibinfo {author} {\bibfnamefont {E.}~\bibnamefont {Greplova}},\ }\bibfield  {title} {\bibinfo {title} {Mode connectivity in neural quantum states},\ }in\ \href {https://openreview.net/forum?id=SC8wiE1QS4} {\emph {\bibinfo {booktitle} {Northern Lights Deep Learning Conference Abstracts 2026}}}\ (\bibinfo {year} {2025})\BibitemShut {NoStop}%
\end{thebibliography}%

\clearpage
\onecolumngrid
\appendix

\setcounter{figure}{0}
\setcounter{table}{0}
\setcounter{equation}{0}
\renewcommand{\thefigure}{S\arabic{figure}}
\renewcommand{\thetable}{S\arabic{table}}
\renewcommand{\theequation}{S\arabic{equation}}

\begin{center}
    \textbf{\large Supplemental Material}
\end{center}

\section{S1. Architecture Details}
Compared to~\cite{vonglehn2023psiformer}, we use pre-norm rather than post-norm, enable QKV biases, and enable orbital biases.

\begin{table}[!htbp]
\centering
\label{tab:sm_architecture}
\begin{tabular}{lc}
\toprule
\textbf{Parameter} & \textbf{Value} \\
\midrule
Transformer Layers & 4 \\
Attention Heads & 4 \\
Model Dimension & 256 \\
MLP Hidden Dimension & 256 \\
Generalized Determinants & 1 \\
Total Parameters & 1,586,182 \\
\bottomrule
\end{tabular}
\caption{Architecture Configuration.}
\end{table}

\section{S2. Training Details}

We use a linear warmup for the learning rate schedule, after which we hold it constant. We do not use momentum or weight decay. In our parameterization regime, SR's cost scales empirically with the batch size squared due to the cost of forming the matrix $\mathbf{X}\mathbf{X}^T$; the formally $\order{M^3}$ inversion has a small constant and was not dominant. To ameliorate this cost, we use subsampling/minibatching: the full batch of $1024$ walkers is split into $4$ chunks of $256$, the SR update is calculated independently within each chunk, and then the chunk updates are averaged. This procedure reduces dominant cost by a factor of $4$ while maintaining training stability.

For Monte Carlo sampling, we use the Metropolis-Hastings algorithm with an isotropic Gaussian proposal distribution moving all coordinates. The proposal width is adjusted adaptively to maintain an acceptance rate of approximately 23.4\%. Adaptation is performed every $10$ iterations: the width is multiplied (divided) by 1.05 if the acceptance probability is too high (low). When estimating the gradient, local energy clipping is applied using a robust median-based estimator: $E_L^{\text{clip}} = \text{median}(E_L) \pm 5 \cdot \text{MAD}(E_L)$, where $\text{MAD}$ is the median absolute deviation.

\begin{table}[!htbp]
\centering
\label{tab:sm_training}
\begin{tabular}{lc}
\toprule
\textbf{Parameter} & \textbf{Value} \\
\midrule
Total Iterations & 50,000 \\
Warmup Iterations & 5,000 \\
Peak Learning Rate & 0.25 \\
SR Diagonal Shift & 0.001 \\
Batch Size & 1,024 \\
SR Subsampling & 4 \\
MCMC Steps per Iteration & 50 \\
Burn-in iterations & 1000 \\
\bottomrule
\end{tabular}
\caption{Training Hyperparameters.}
\end{table}

\begin{table}[!htbp]
\centering
\label{tab:sm_autoneb}
\begin{tabular}{lc}
\toprule
\textbf{Parameter} & \textbf{Value} \\
\midrule
Total Iterations & 5,000 \\
Warmup Iterations & 250 \\
Initial Pivots & 5 \\
Insertion Interval & 500 iterations \\
Spring Constant ($k$) & 0.0 \\
Batch Size & 512 \\
SR Subsampling & 2 \\
Burn-in iterations & 500 \\
\bottomrule
\end{tabular}
\caption{GeoNEB Hyperparameters.}
\end{table}

\end{document}